\documentclass[10pt]{iopart}

\usepackage{iopams}  
\usepackage{graphicx}
\usepackage{hyperref}
\usepackage{comment}
\usepackage{esint}
\usepackage{mwe}
\usepackage[table]{xcolor}
\usepackage{tikz}
\usetikzlibrary{shapes.geometric}

\hypersetup{
    colorlinks=true,
    linkcolor=green,   
    citecolor=green,   
    urlcolor=green
}

\newcommand{\bm}[1]{\mathbf{#1}}
\newcommand{\fs}[1]{\left< #1\right>_\psi}

\newcommand{\avgt}[1]{\left<#1 \right>_t}

\newcommand{\eqref}[1]{Eq.~\eref{#1}}

\definecolor{tabblue}{rgb}{0.121, 0.466, 0.705}
\definecolor{taborange}{rgb}{1.000, 0.498, 0.055}
\definecolor{tabgreen}{rgb}{0.173, 0.627, 0.173}
\definecolor{tabred}{rgb}{0.839, 0.153, 0.157}
\definecolor{tabpurple}{rgb}{0.580, 0.404, 0.741}
\definecolor{tabbrown}{rgb}{0.549, 0.337, 0.294}
\definecolor{tabpink}{rgb}{0.890, 0.466, 0.760}
\definecolor{tabgray}{rgb}{0.498, 0.498, 0.498}
\definecolor{tabolive}{rgb}{0.737, 0.741, 0.133}
\definecolor{tabcyan}{rgb}{0.090, 0.745, 0.811}

\definecolor{color2p7}{rgb}{0.455, 0.753, 0.988}
\definecolor{color3p8}{rgb}{0.227, 0.525, 1.000}
\definecolor{color4p5}{rgb}{0.122, 0.467, 0.706}
\definecolor{color4p8}{rgb}{0.043, 0.235, 0.365}

\newcommand{\markerCircle}[1][black]{\tikz[baseline=-0.5ex]\draw[fill=#1,draw=#1] (0,0) circle (0.6ex);}
\newcommand{\markerSquare}[1][black]{\tikz[baseline=-0.5ex]\draw[fill=#1,draw=#1] (0,0) rectangle (0.12,0.12);}
\newcommand{\markerDiamond}[1][black]{\tikz[baseline=-0.5ex]\draw[fill=#1,draw=#1,rotate=45] (0,0) rectangle (0.12,0.12);}
\newcommand{\markerStar}[1][black]{\tikz[baseline=-0.5ex, scale=0.14]\node[star,star points=5,star point ratio=2.25,fill=#1,inner sep=1pt,draw=#1] at (0,0) {};}

\newcommand{\markerCross}[1][black]{\tikz[baseline=-0.5ex]\draw[#1, line width=0.5pt,scale=1.3] (-0.08,-0.08) -- (0.08,0.08) (-0.08,0.08) -- (0.08,-0.08);}

\newcommand{\markerRightTriangle}[1][black]{%
  \tikz[baseline=-0.5ex,scale=0.75]%
    \draw[fill=#1,draw=#1] (0,0.14) -- (0.24,0) -- (0,-0.14) -- cycle;%
}
\begin{document}

\title[Validation of pre L-H transition]{Pre L-H Transition Radial Electric Field and Transport Validations of Edge and Scrape-off Layer Gyrokinetic Simulations at ASDEX Upgrade}

\author{B. J. Frei$^1$, C. Angioni$^1$, G. Lo-Cascio$^1$, W. Zholobenko$^1$, P. Ulbl$^1$, R. Bilato$^1$, F. Jenko$^{1}$, the ASDEX Upgrade Team$^{2}$}

\address{$^1$ Max-Planck Institute for Plasma Physics, Boltzmannstr. 2, Garching, D-85748, Germany}
\address{$^{2}$ See author list of H. Zohm et al 2024 Nucl. Fusion https://doi.org/10.1088/1741-4326/ad249d}

\ead{baptiste.frei@ipp.mpg.de}
\vspace{10pt}
\begin{indented}
\item[]March 2025
\end{indented}

\begin{abstract}
This work presents a stepwise validation of the evolution of the radial electric field $E_r$ and heat transport during the pre L-H transition phase using full-$f$ gyrokinetic simulations of the edge and scrape-off layer in the ASDEX Upgrade (AUG) tokamak, including X-point geometry. Several L-mode time slices up to the L-H transition from a dedicated hydrogen discharge, featuring stepwise increases in ECRH input power, are selected [N. Bonanomi \textit{et al.}, Phys. Plasmas 31, 072302 (2024)] and simulated with the \texttt{GENE-X} code. As the edge boundary conditions are progressively increased between the time slices, particle and heat fluxes rise, and the radial electric field $E_r$ well deepens. A detailed validation of the $E_r$ profiles and of the $E_r$ well depth shows excellent agreement with experimental measurements at the successive time slices approaching the L-H transition. A force balance decomposition identifies turbulence-driven poloidal flows as the dominant contribution within the $E_r$ well. Edge turbulence is governed by a competition between electron drift waves and trapped-electron modes. The introduction of an edge density source, modeling neutral gas ionization, is shown to be essential to reproduce experimentally relevant density profiles, $E_r$, and edge ion heat fluxes, which are dominated by both turbulent and diamagnetic contributions. This stepwise validation constitutes an important milestone toward predictive, first-principles gyrokinetic simulations of the L-H transition power threshold.
\end{abstract}

%
%
%
\ioptwocol

\section{Introduction}
\label{sec:sec1}
Accurate prediction of turbulence and transport in the plasma edge and scrape-off layer (SOL) is a critical challenge for future fusion reactors, as these coupled regions directly determine the energy confinement time and the achievable fusion power \cite{shimada2007,goldston2011}. Since both quantities depend sensitively on the edge pressure, operation in the high-confinement (H-mode) regime—characterized by the formation of an edge pressure pedestal—remains highly attractive \cite{wagner1982,zohm1996,doyle2007}. It is now widely accepted that the transition from low- to high-confinement (L–H transition) is governed by edge turbulence and transport processes \cite{diamond2005}. Experiments at ASDEX Upgrade (AUG) have demonstrated a strong correlation between the H-mode power threshold $P_{\mathrm{LH}}$ and thresholds in the edge ion heat flux $Q_i$ \cite{ryter2014,ryter2015}, as well as in the minimum of the radial electric field $E_r$ and its shear \cite{sauter2011,schmitz2012}. However, the strong coupling between the confined plasma edge and the SOL, together with the presence of magnetic X-points, challenges the applicability of local transport models, which are unable to capture the nonlinear interplay between turbulence, profiles, and flows in this region.

A recent validation study based on local and $\delta f$ gyrokinetic (GK) simulations using the \texttt{GENE} code \cite{jenko2000} investigated a dedicated hydrogen discharge at AUG just prior to the L–H transition, providing an important assessment of the local turbulent transport paradigm under extreme edge conditions \cite{bonanomi2024}. It is found that when all simulation inputs were taken directly from experimental profiles, the predicted ion and electron turbulent heat fluxes are consistent with experiments across different L-mode time slices approaching the L–H transition. The combined destabilization of long wavelength modes by electromagnetic effects and stabilization by $\bm{E} \times \bm{B}$ shear was identified as the key mechanism preventing a sharp increase in turbulent transport as the edge pressure steepens. Despite these promising results, the predictive capability of local approaches remains fundamentally limited: they rely on experimentally prescribed equilibrium profiles and $E_r$ as inputs and, therefore, cannot provide a predictive description of edge dynamics. This limitation is particularly critical given the central role of $E_r$ (and its shear) in regulating edge turbulence and transport (e.g., $Q_i$) approaching the L–H transition and, therefore, the total amount of power required to enter in H-mode.

In this context, the present work goes beyond the limitations of local turbulence modelling by employing first-principles, electromagnetic, full-$f$, and collisional GK simulations of edge and SOL turbulence including X-point geometry. Within the full-$f$ GK framework, a self-consistent description of edge and SOL is achieved. We therefore extend the work of \cite{bonanomi2024} to global edge and SOL turbulence modelling using the \texttt{GENE-X} code \cite{michels2022,frei2025vspec}. More precisely, we consider the same dedicated AUG hydrogen discharge at different L-mode time slices approaching the experimental L-H transition. Rather than attempting to simulate the slow temporal evolution of plasma pressure profiles under prescribed input power \cite{zholobenko2026}, we adopt a step-wise validation strategy based on independent simulations of successive experimental time slices. In this approach, the inner edge boundary is fixed to the experimental profiles, while the edge and SOL profiles evolve self-consistently in response to turbulence, transport and density source. Accordingly, we emphasize that all references to the \textit{L-H transition} in this work refer to the \textit{experimental L-H transition}.

The present analysis reveals the key role of turbulence-driven poloidal flows in the formation and deepening of the $E_r$ well in agreement with experimental measurements as the L–H transition is approached. In addition, the inclusion of an edge density source is shown to be essential for reproducing experimentally compatible ion and electron heat fluxes, $E_r$, and density profile close to the transition. This validation demonstrates the maturity of global edge–SOL turbulence simulations in reproducing critical L-mode physics ingredients, such as $E_r$ and $Q_i$, that are believed to be essential for triggering the L–H transition, thereby representing an important step towards predictive simulations for reactor-relevant conditions. 

The paper is structured as follows. First, we describe the dedicated AUG hydrogen discharge considered in this study. The full-$f$ edge and SOL GK framework implemented in \texttt{GENE-X}, including the treatment of density sources and the numerical setup, is then introduced. Particle and energy balance equations are established and verified to assess the quasi-steady state conditions in the presence of a density source. Outboard midplane (OMP) profiles are validated against experimental measurements, with particular emphasis on the radial electric field. A force balance analysis is then used to investigate the role of poloidal flows in the evolution of the $E_r$ well as the L–H transition is approached. Edge turbulence is characterised, and the predicted ion and electron heat transports are compared with the results of \cite{bonanomi2024}. Finally, the main findings and conclusions are summarized.

\section{Hydrogen Experiments at ASDEX Upgrade}
\label{sec:sec2}

\begin{figure}[h]
    \centering
    \includegraphics[scale=0.55]{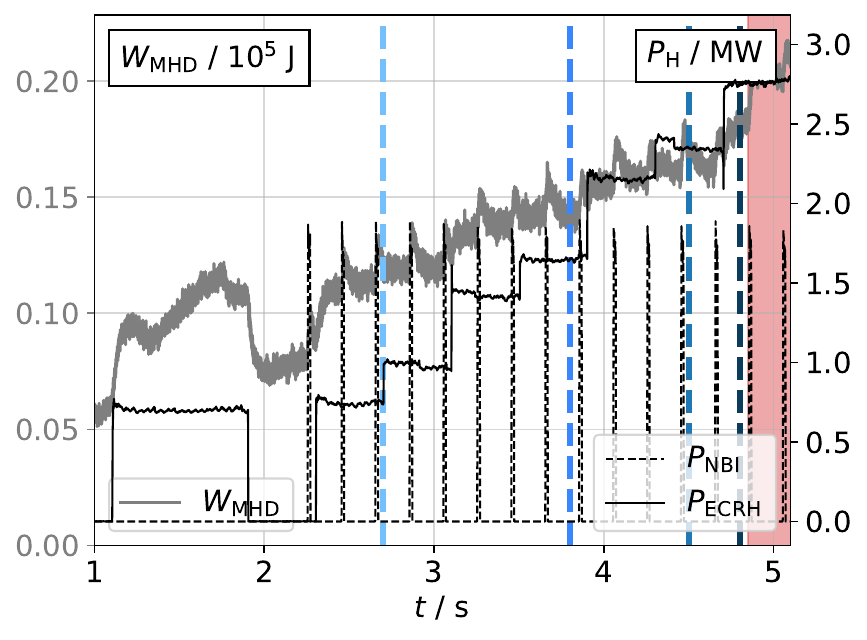}
    \caption{Experimental evolution of the plasma stored energy $W_{\textrm{MHD}}$ (gray solid line). The time traces of the experimental input heating power, $P_{\textrm{H}}$, composed of the ECRH power (solid black line) and NBI blips (dashed black lines), are also shown on the right $y$-axis. Vertical colored dashed lines (see \ref{table:simsummary}) indicate the selected L-mode time slices. The red shaded area marks the H-mode phase for $t \gtrsim 4.85$~s. Colors}
    \label{fig:fig1}
\end{figure}

The plasma discharge ($\#38176$) considered in this work is a dedicated hydrogen (H) experiment performed at AUG (major radius $R = 1.65$~m, minor radius $a = 0.5$~m, and a full tungsten wall). This discharge is the same as the one previously considered in \cite{bonanomi2024}. The plasma scenario is characterized by a magnetic field of $B = 2.5$~T in a favourable lower single-null configuration. The core line-averaged electron density is $\left< n_e\right> \simeq 3 \times 10^{19}$~m$^{-3}$ (lower density branch) and the plasma current is $I_p = 1.2$~MA ($q_{95} \simeq 3.6$). The time evolution of the plasma stored energy is shown in \fref{fig:fig1}. In this discharge, the electron cyclotron resonance heating (ECRH) power ($P_{\textrm{ECRH}}$) is progressively increased by $300$~kW every $400$~ms. Neutral beam injection (NBI) blips to allow ion temperature measurements, with a peak power of $P_{\mathrm{NBI}} = 2.5$~MW, are applied every $200$~ms for $16$~ms. The time traces of the ECRH and NBI input powers are also shown in \fref{fig:fig1}.

The electron density ($n_e$) profiles were obtained from Thomson scattering measurements, interferometry, and Li-Beam emission diagnostics \cite{willensdorfer2012}. The electron temperature ($T_e$) profiles were measured using Electron Cyclotron Emission (ECE) diagnostics as well as Thomson scattering (TS) systems. On the other hand, the ion temperature ($T_i$) measurements were performed through charge exchange (CX) \cite{cavedon2017} using the NBI blips. Doppler reflectometry was applied to infer the radial profile of $E_r$ in the edge. This technique measures the Doppler frequency shift induced by fluctuating structures moving relative to the laboratory frame, capturing contributions from both the plasma $\bm E \times \bm B$ velocity and the turbulence phase velocity in the $\bm E \times \bm B$ frame. Averages between different times slices have been performed to obtain $E_r$ profiles at the considered time slices \cite{bonanomi2024}. Finally, to study the evolution of the $E_r$ minimum, a diagnostic based on active spectroscopy of helium II lines was employed \cite{plank2023}. 

In this discharge, the L-H transition occurs at $t \simeq 4.85$~s and is trigger by a NBI blip where the ECRH heating power reaches $2.75$~MW. The experimental net input power ($P_{\mathrm{net}}$) is calculated as $P_{\mathrm{net}} = P_{\mathrm{H}} - dW_{\mathrm{MHD}}/dt$, where $P_{\mathrm{H}}$ denotes the total (ECRH, NBI and Ohmic) heating power. Note that the radiated power is not removed. The experimental L-H power threshold, denoted by $P_{\mathrm{LH}}$, is defined as $P_{\mathrm{net}}$ at the time of the L-H transition, i.e. at $P_{\mathrm{LH}} = P_{\mathrm{net}}$ at $t \simeq 4.85$~s.

Similar to \cite{bonanomi2024}, we analyze the time slices, $t = 2.7$~s, $3.8$~s, $4.5$~s, and $4.8$~s, in L-mode prior to the L-H transition ($t \lesssim 4.85$~s). These time slices (indicated in \fref{fig:fig1}) correspond to progressively increasing experimental $P_{\mathrm{net}}$, such that (relative to the L-H transition threshold $P_{\mathrm{LH}}$) $P_{\mathrm{net}} / P_{\mathrm{LH}} = 0.5$, $0.8$, $0.95$, and $0.99$, respectively. We perform a single and independent \texttt{GENE-X} simulation for each time slice.

Finally, we clarify that, in the present work, references to \textit{the L-H transition} relate to the \textit{experimental L-H transition} observed at $t \simeq 4.85$~s (see \fref{fig:fig1}) since the simulations presented here are conducted with fixed input power (provided by the Dirichlet boundary conditions). Therefore, we do \textit{not} model the dynamics of an actual L-H transition \cite{zholobenko2026}.

\section{Gyrokinetic Edge and SOL Turbulence Modelling}
\label{sec:sec3}

To model self-consistently the edge and SOL turbulence, we use the full-$f$ GK \texttt{GENE-X} code \cite{michels2021} designed for edge and SOL turbulence with magnetic X-points. More precisely, the \texttt{GENE-X} code solves for the full (gyroaveraged part) of the \textit{gyrocenter} distribution function, $ f_\alpha = f_\alpha(\bm{R}, v_\parallel, \mu,t)$, for particle species $\alpha$ (mass $m_\alpha$ and charge $q_\alpha$) in the gyrocenter phase-space described by the coordinates $(\bm{R}, v_\parallel, \mu, \theta)$. Here, $\bm{R}$ is the gyrocenter position, $v_\parallel = \bm{b} \cdot \bm{v}$ is the velocity parallel to the equilibrium magnetic field, and $\mu = m_\alpha v_\perp^2 / (2B)$ is the magnetic moment. The unit vector $\bm{b} = \bm{B} / B$ represents the direction of the equilibrium magnetic field $\bm{B}$, while $v_\perp = \left|\bm{v} - \bm{b} v_\parallel\right|$ is the velocity component perpendicular to $\bm{B}$. The self-consistent evolution of $f_\alpha$ is obtained from the long-wavelength electromagnetic and collisional GK Vlasov equation. For better computational efficiency, we use the spectral implementation of \texttt{GENE-X} in the present work \cite{frei2025vspec}. Within the spectral approach, $f_\alpha$ is expanded onto a set of orthogonal and scaled Hermite and Laguerre polynomials, such that $f_\alpha$ is approximated by 

\begin{equation} \label{eq:faspectral}
f_\alpha \simeq \sum_{p =0}^{N_{v_\parallel}-1} \sum_{j =0}^{N_\mu-1} \mathcal{N}_\alpha^{pj} \hat H_p(\hat v_{\parallel \alpha}) L_j(\hat \mu_\alpha) F_{\mathcal{M} \alpha}.
\end{equation}
Here, $\hat H_p(\hat v_{\parallel \alpha}) = H_p(\hat v_{\parallel \alpha})  / \sqrt{2^p p !} $ and $L_j(\hat \mu_\alpha)$ are the Hermite and Laguerre polynomials respectively \cite{gradshteyn2014}, with the scaled velocity-space coordinates $\hat v_{\parallel \alpha} = v_\parallel \sqrt{2 \tau_\alpha / m_\alpha}$ and $\hat \mu_\alpha = \mu B  / \tau_\alpha$ as arguments. $\tau_\alpha$ is a constant reference temperature introduced to adjust the velocity-space basis \cite{frei2025vspec}. In \eqref{eq:faspectral}, the spectral coefficients, $\mathcal{N}_\alpha^{pj}$, are defined as Hermite-Laguerre weighted moments of $f_\alpha$,

\begin{equation} \label{eq:npj}
  \mathcal{N}_\alpha^{pj}  = \int d W  \hat H_p(\hat v_{\parallel \alpha}) L_j(\hat \mu_\alpha) f_\alpha.
\end{equation}
and are related to fluid quantities \cite{frei2025vspec}. In the spectral formulation, the dynamics is entirely described by the spectral coefficients, $\mathcal{N}_\alpha^{pj}$, which are evolved according to 

\begin{eqnarray} \label{eq:spectralvlasov}
    & \frac{\partial}{\partial t } \mathcal{N}^{pj}_\alpha  + \nabla  \cdot \bm \Gamma_\alpha^{pj} + \mathcal{F}_{\alpha \ell k}^{pj}\mathcal{N}^{\ell k}_\alpha + \mathcal{D}_{\alpha \ell k}^{pj}\mathcal{N}^{\ell k}_\alpha \frac{\partial A_{1 \parallel} }{\partial t}  \nonumber \\
    & = \sum_{\beta} \mathcal{C}_{\alpha \beta}^{pj} + \mathcal{S}_{n_\alpha}^{pj}.
\end{eqnarray}
On the right-hand side of \eqref{eq:spectralvlasov}, collisional effects ($\mathcal{C}_{\alpha \beta}^{pj}$) are modeled using a full-$f$ Lernard-Bernstein Daugherty collision operator \cite{ulbl2022,frei2025vspec}. The analytical expressions of $\bm \Gamma_\alpha^{pj}$, $\mathcal{F}_{\alpha \ell k}^{pj}$, $\mathcal{D}_{\alpha \ell k}^{pj}$, and $\mathcal{C}_{\alpha \beta}^{pj}$ can be found in \cite{frei2025vspec}. The term $\mathcal{S}_{n_\alpha}^{pj}$ is the spectral representation associated with a localized edge density source, which we introduce in \sref{subsec3.1}. To obtain the evolution of the electrostatic potential, $\phi_1$, and parallel vector potential $A_{\parallel 1}$, the spectral quasi-neutrality condition, Ampere's and Ohm's law, are used. These equations are reported in \cite{frei2025vspec}. In the present work, a simple closure by truncation is applied to \eqref{eq:spectralvlasov} and a finite number of $N_{v_\parallel}$ (Hermite) and $N_{\mu}$ (Laguerre) spectral coefficients are retained.

\subsection{Edge Density Source}
\label{subsec3.1}

While an accurate description of edge density fueling requires a self-consistent treatment of neutral dynamics \cite{zholobenko2021}, we consider a \textit{ad-hoc} kinetic density source model $\mathcal{S}_{n_\alpha}$, which is introduced in the GK Vlasov equation to mimic the presence of an ionization source. The density source model, $\mathcal{S}_{n_\alpha}$, is defined by \cite{sarazin2010}

\begin{eqnarray} \label{eq:source}
\mathcal{S}_{n_\alpha} & = \frac{\mathbb{S}_{n_\alpha}}{\left(2 \pi \right)^{3/2} (T_{S \alpha} / m_\alpha)^{3/2}} \nonumber \\
& \times \left[ 1 - \frac{\hat{H}_{2} (\hat v_{\parallel S})}{\sqrt{2}}  +  L_{1} (\hat \mu_{ S})\right]  e^{-\hat v_{\parallel S}^2 - \hat \mu_{ S}},
\end{eqnarray}
with $\hat v_{\parallel S} = v_\parallel  / \sqrt{2 T_{S \alpha} / m_\alpha}$ and $\hat \mu_{ S} = \mu B / T_{S \alpha}$. Here, $T_{S \alpha}$ is the source reference temperature, which is assumed constant and $\mathbb{S}_{n_\alpha} = \int d W \mathcal{S}_{n_\alpha}$ is the \textit{fluid} density source (units of m$^{-3}$ s$^{-1}$). The analytical form of $\mathcal{S}_{n_\alpha}$ stems from the fact that we require that the fluid energy source $\mathbb{S}_{\epsilon \alpha}  = \int d W m_\alpha v^2 \mathcal{S}_{n_\alpha} / 2$ associated with $\mathcal{S}_{n_\alpha}$, vanishes such that $\mathbb{S}_{\epsilon \alpha} = 0$. This condition is ensured by the presence of the $\hat{H}_{2} (\hat v_{\parallel S})$ and $L_1(\hat \mu_{ S})$ terms in \eqref{eq:source}. For simplicity, we assume that $\mathbb{S}_{n_\alpha}$ to be poloidally symmetric and axisymmetric, while its radial dependence is given by a Gaussian shape function (center $\rho_S$ and width $L_S$), such that $\mathbb{S}_{n_\alpha} = C_n  S_{n_\alpha} e^{ - (\rho - \rho_S) / L_S}$ with $C_n = (\int d V e^{ - (\rho - \rho_S) / L_S})^{-1}$. We also require that the density source does not inject vorticity, such that $\sum_\alpha q_\alpha S_{n_\alpha} = 0$. As we consider an hydrogen plasma, we assume $\mathcal{S}_{n_e} = \mathcal{S}_{n_i} = \mathcal{S}_{n} $ and $S_{n_e} = S_{n_i} = S_n$. We note that, since the long-wavelength approximation is employed in our model, polarization effects in the density source are neglected, i.e. we do not distinguish between a \textit{particle} and \textit{gyrocenter} density source. 

The spectral representation of $\mathcal{S}_{n_\alpha}$ is obtained by projecting \eqref{eq:source} onto the scaled Hermite-Laguerre basis \cite{frei2025vspec},

 \begin{equation} \label{eq:sourcevspec}
 \mathcal{S}_{ \alpha}^{pj} = \mathbb{S}_{n_\alpha}    \left[  \delta_p^0 \delta_j^0 - \frac{1}{\sqrt{2}} \delta_p^2 \delta_j^0  +  \delta_p^0 \delta_j^1 \right],
\end{equation}
 assuming $\tau_\alpha = T_{S \alpha}$ for simplicity. Note that the spectral expansion given in \eqref{eq:sourcevspec} can be generalized to the case $\tau_\alpha \neq T_{S \alpha}$ and will be reported in a future publication. By construction, the amplitude, the radial position and the width of the density source model are parameters that need to be prescribed. In the present work, these parameters are adjusted to match the experimental density profile close to the separatrix near the L-H transition. While we consider a single source model, the source density parameters are varied in \ref{sec:appendixb} to assess their effects on the present results.

\subsection{Numerical Setup}
\label{subsec3.2}

We simulate in total four distinct time slices (shown in \fref{fig:fig1}), each corresponding to a different \textit{experimental} $ P_{\mathrm{net}} / P_{\mathrm{LH}} $ value approaching the L-H transition. To investigate the effects of edge particle fueling, we also perform an additional simulation at $t = 4.8$~s ($ P_{\mathrm{net}} / P_{\mathrm{LH}} = 0.99$) by introducing a localized density source, $S_{n}$, near the separatrix. The simulations corresponding to other time slices do not include a density source $S_n$, since we focus on the effect of $S_n$ prior to the L-H transition. The density source, $S_n$, is centered around $\rho_S = 0.99$ with width $L_S = 0.005$. The amplitude is manually adjusted to match the experimental density profile near the separatrix. It is found that a constant amplitude of $S_{n} = 1.6 \times 10^{22}$~s$^{-1}$ was sufficient in this case. A summary of the simulations performed in this work is shown in \tref{table:simsummary}. For consistency, we use the labels and associated colors defined in \fref{fig:fig1}.  

\begin{table*}[t]
\centering
\begin{tabular}{cccccccc}
$t$ / s & $P_{\mathrm{net}} / P_{\mathrm{LH}}$ &  $n_e$ / m$^{-3}$ & $T_{e}$ / keV & $T_i$ / keV &w/ $S_{n}$ &  Color \\
\hline
\hline
$2.7$ & $0.5$ & $2.5$ & $0.5$ & $0.4$  & No  &  \textcolor{color2p7}{\rule{2ex}{1.5ex}} \\
$3.8$ & $0.8$ & $2.6$ & $0.6$ &  $0.425$ & No  &  \textcolor{color3p8}{\rule{2ex}{1.5ex}} \\
$4.5$ & $0.9$ & $2.75$ & $0.7$ & $0.475$&No  &  \textcolor{color4p5}{\rule{2ex}{1.5ex}} \\
$4.8$ & $0.99$& $2.75$ & $0.75$ & $0.5$ &No  &  \textcolor{color4p8}{\rule{2ex}{1.5ex}} \\
$4.8$ & $0.99$& $2.75$ & $0.75$ & $0.5$ & Yes &  \textcolor{tabred}{\rule{2ex}{1.5ex}} \\
\end{tabular}
\caption{Summary of \texttt{GENE-X} simulations: time slices (see \fref{fig:fig1}), experimental $P_{\mathrm{net}} / P_{\mathrm{LH}}$ values \cite{bonanomi2024}, experimental density ($n_e$) and electron ($T_e$) and ion ($T_i$) temperatures used as inner boundary conditions imposed at $\rho_{\mathrm{pol}} = 0.9$, edge density source (w/ $S_n$), and colors considered throughout the manuscript.}
\label{table:simsummary}
\end{table*}

To minimize variations in magnetic equilibrium, we use the same magnetic configuration for all simulations, based on the one reconstructed at $t = 4.8$~s.

Boundary conditions for the spectral coefficients and electromagnetic fields are required. We adopt here Dirichlet boundary conditions in the SOL. At the divertor targets and external boundary of the numerical domain ($\rho_{\mathrm{pol}} = 1.04$), the spectral coefficients are fixed to their initial Maxwellian values, evaluated using the SOL experimental densities and temperatures \cite{frei2025vspec}. More precisely, the density and (electron and ion) temperatures are fixed to $2 \times 10^{18}$~m$^{-3}$ and $35$~eV, respectively at the SOL boundary located at $\rho_{\mathrm{pol}} = 1.04$. The electromagnetic fields, $\phi_1$ and $A_{\parallel 1}$, are set to zero at all boundaries. For the edge boundary conditions at $\rho_{\mathrm{pol}} = 0.9$, we also apply Dirichlet conditions using a local spectral Maxwellian distribution function \cite{frei2025vspec}, with density and temperature calibrated to the experimental measurements, as reported in \tref{table:simsummary}. On the other hand, for the spectral coefficients with $p$ odd (associated with fluxes), we impose zero zonal Neumann conditions. This allows flux variables to develop a finite value near the edge boundary. Applying Neumann conditions at the inner edge appears to be important to prevent spurious flow shear layers from establishing \cite{frei2025}. Although the profiles are fixed to values close to experimental measurements at the edge and SOL boundaries, they are free to evolve self-consistently between $\rho_{\mathrm{pol}} = 0.9$ and $1.04$

Energy and particles (without edge density source) are injected in the simulations through the inner edge boundary via the combination of Dirichlet boundary conditions and spatial hyperdiffusion applied within buffer region near the boundary in the poloidal plane. This numerical approach is similar to an adaptive Krook operator, driving the distribution function to a Maxwellian distribution at the boundary. Therefore, the increase of input power across simulations (i.e. time slices) results from the increase of the edge boundary values between each simulation (see \tref{table:simsummary}). Particle and energy losses in the SOL are provided by a buffer region located near the boundary of the numerical domain. An additional particle source is provided by $S_n$ near inside the separatrix.

\begin{figure}[h]
    \centering
    \includegraphics[scale=0.55]{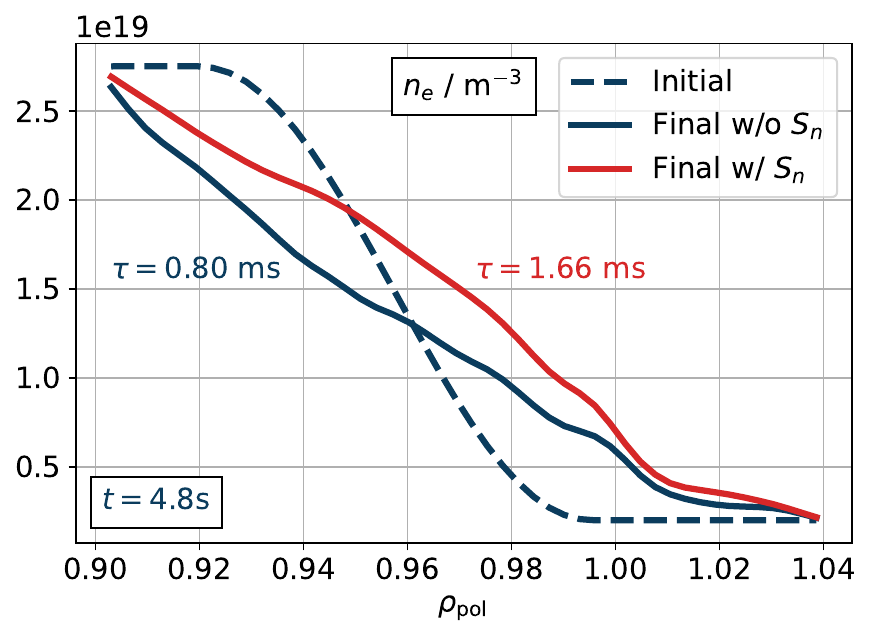}
    \caption{Initial (dashed) and quasi-steady state (solid) toroidally averaged OMP electron density ($n_e$) profiles, at $t=4.8$~s, without (dark blue) and with (red) density source $S_n$ (see \tref{table:simsummary}).}
    \label{fig:fig2}
\end{figure}

To start the simulations, the spectral coefficients are initialized using a local spectral Maxwellian distribution function \cite{frei2025vspec}. The corresponding initial density and temperature profiles are prescribed using smooth, axisymmetric analytical $\sin$-shaped functions~\cite{michels2022}. The initial electron density profile used for the simulations at $t = 4.8$~s is shown in \fref{fig:fig2}. Identical analytical profiles are used in all simulations, except at the inner boundary, where the values are adjusted to match those reported in \tref{table:simsummary}. The electrostatic potential ($\phi_1$) and perturbed parallel magnetic potential ($A_{\parallel 1}$) fields are initially set to zero everywhere. In \sref{sec:appendixa}, we detail the normalization, numerical resolution and the overall computational cost used in the present work.

\section{Quasi-Steady State Particle and Energy Balance}
\label{sec:sec3}

\begin{figure}[h]
    \centering
    \includegraphics[scale=0.55]{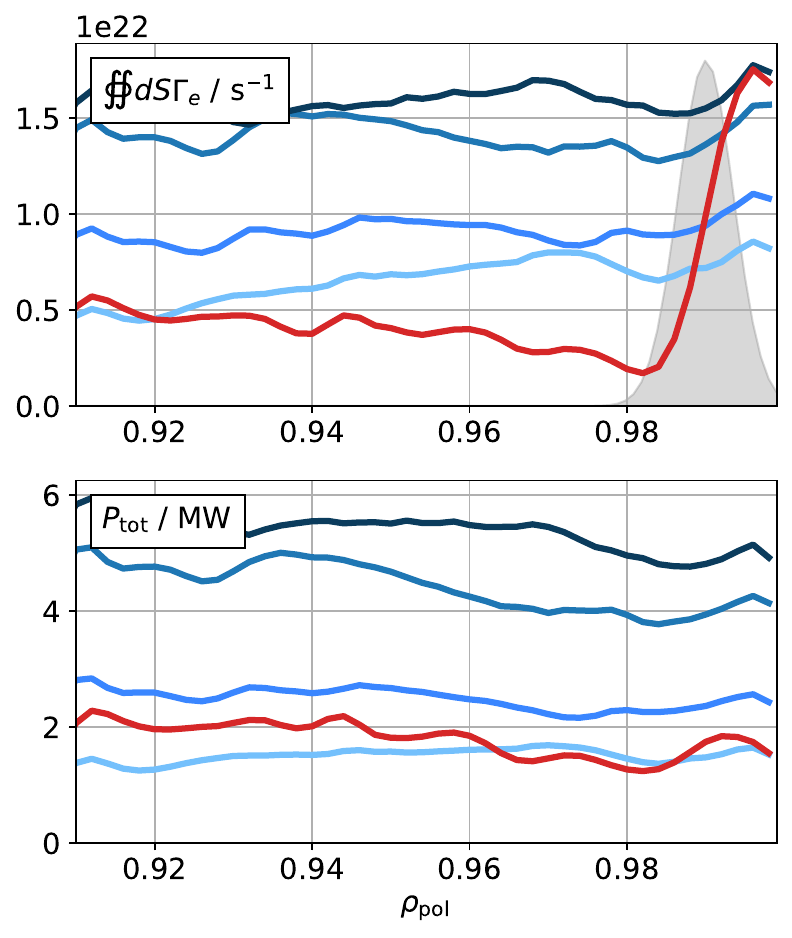}
    \caption{Total surface integrated radial (top) particle $\Gamma_{e}$ and (bottom) heat $Q_{tot} = \sum_\alpha Q_{\alpha} $ fluxes as a function of $\rho_\mathrm{pol}$ obtained in the different simulations at quasi-steady-state. The fluxes are averaged over $0.1$~ms. The localized particle source, $\mathbb{S}_n$, is also shown and scaled such that its maximum equals $S_{n} = 1.6 \times 10^{22}$~s$^{-1}$. }
    \label{fig:fig3}
\end{figure}

At quasi-steady state (QSS), particle and heat fluxes must establish in order to balance the presence of volumetic sources and sinks. In the present case, the particle and energy sources are provided by the inner edge boundary and by the localized source, if $S_n$ is added. On the other hand, the particle and energy losses are provided by the SOL. In this section, we aim to qualitatively assess the QSS by considering the particle and (total) energy balance equations that we derive from the GK Vlasov equation.

The balance equation for the particle (or \textit{gyrocenter}) $n_\alpha$ is obtained from the zeroth order moment of the GK Vlasov equation, \eqref{eq:spectralvlasov}. Applying then the flux surface average (FSA) operator, $\fs{A} =  (V')^{-1} \oint d S A / \left|\nabla \psi \right|$ ($V(\psi)$ is the volume enclosed by the $\psi$ flux surface and $V' = d V(\psi) / d \psi$), we derive

\begin{equation} \label{eq:densitybalance}
\partial_t \fs{n_\alpha} + \frac{1}{V^\prime} \frac{d}{d \psi} \left(V'\fs{\bm{\Gamma}_\alpha \cdot \nabla \psi } \right)=   \fs{\mathbb{S}_{n_\alpha}},
\end{equation}
where $\bm{\Gamma}_\alpha = \int dW f_\alpha \dot{\bm{R}}$ is the \textit{total gyrocenter} flux containing both the electrostatic and electromagnetic turbulent and diamagnetic (or neoclassical) contributions. At QSS, the time derivative in \eqref{eq:densitybalance} vanishes when averaging over time. Thus, applying the time-average operator, $\left< \cdot \right>_t = \int_{t}^{t + \Delta t} d \tau \cdot / \Delta t$ (with $\Delta t \sim 0.1$~ms) at QSS, we obtain

\begin{equation} \label{eq:densitybalance2}
 \frac{1}{V^\prime} \frac{d}{d \psi} \left(V'\fs{\bm{\Gamma}_\alpha \cdot \nabla \psi } \right) =   \fs{\mathbb{S}_{n_\alpha}},
\end{equation}
omitting the $\avgt{\cdot}$ notation for simplicity.
Integrating \eqref{eq:densitybalance2} from the edge ($\rho_{\mathrm{pol}} \simeq 0.9$) and to the separatrix ($\rho_{\mathrm{pol}} \simeq 0.99$), yields

\begin{equation} \label{eq:densitybalanceintegrated}
\left.V' \fs{ \bm{\Gamma}_\alpha  \cdot  \nabla \psi}\right|_{sep} =\left. V' \fs{ \bm{\Gamma}_\alpha  \cdot  \nabla \psi}\right|_{edge}   + S_{n_\alpha}.
\end{equation}
From \eqref{eq:densitybalanceintegrated}, the particle flux increases towards the separatrix due to the presence of the localized particle source $S_n$. Note that, as a consequence of the Dirichlet boundary conditions imposed at $\rho_{\mathrm{pol}} \simeq 0.9$, a finite (outwards) particle flux $\fs{ \bm{\Gamma}_\alpha  \cdot  \nabla \psi}$ appears in our simulations

Similarly to the density balance equation given in \eqref{eq:densitybalance}, an energy balance equation can be derived by evaluating the $(p,j) = (2,0)$ and $(0,1)$ spectral GK Vlasov equations using \eqref{eq:spectralvlasov}. More precisely, the balance equation for the total energy $E_\alpha = \int d W m_\alpha v^2 f_\alpha / 2 =  3 n_\alpha T_\alpha / 2 $ at QSS reads

\begin{eqnarray} \label{eq:nrgbalance}
  & \frac{1}{V^\prime} \frac{d}{d \psi} \left(V'\fs{ \bm Q_{\alpha} \cdot \nabla \psi } \right)   -  \fs{F_{\parallel \alpha}}- \fs{\nabla \ln B \cdot \bm Q_{ \alpha}^\perp} \nonumber \\
  & = \sum_{\beta \neq \alpha} \fs{Q_{\alpha \beta}} + \fs{\mathbb{S}_{\epsilon \alpha}},
\end{eqnarray}
with the \textit{total} \textit{gyrocenter} energy fluxes, $\bm Q_{\alpha} = \int d W m_\alpha v^2 f_\alpha \dot{\bm R} / 2 $ and $\bm Q_{ \alpha}^\perp = \int d W \mu B \dot{\bm R} f_\alpha$. In \eqref{eq:nrgbalance}, the time-average operator $\avgt{\cdot}$ is assumed. $F_{\parallel \alpha} =  \int d W   m_\alpha \dot v_\parallel  v_\parallel f_\alpha$ represents the power associated with the parallel force, $m_\alpha \dot v_\parallel$. The terms on the right-hand side are the collisional heating, $Q_{\alpha \beta} = \int d W m_\alpha v^2 C_{\alpha \beta}(f_\alpha, f_\beta) / 2 $, and fluid energy source, 
$\mathbb{S}_{\epsilon \alpha}$, associated with $\mathcal{S}_{n_\alpha}$, respectively. In the case of the density source detailed in \sref{subsec3.1}, $\mathbb{S}_{\epsilon \alpha}$ vanishes. 

It is worth noticing that the energy balance equation, given in \eqref{eq:nrgbalance}, can be used in combination with the density balance equation, \eqref{eq:densitybalance}, to derive an evolution equation for the total temperature $T_\alpha$. From this evolution equation (not shown here), it can be observed that $\mathbb{S}_{n_\alpha}$ provides an effective local temperature sink of amplitude $- 1.5 T_\alpha \mathbb{S}_{n_\alpha} / n_\alpha$ if $\mathbb{S}_{\epsilon \alpha} = 0$, while $- 1.5 \mathbb{S}_{n_\alpha} ( T_\alpha - T_{S \alpha})/ n_\alpha$ otherwise. 

Summing \eqref{eq:nrgbalance} over all species, the total energy balance equation becomes

\begin{eqnarray} \label{eq:steadystateheat}
   & \frac{1}{V^\prime} \frac{d}{d \psi} \left(V' \fs{\bm Q \cdot \nabla \psi}   \right)  -  \sum_\alpha \fs{F_{\parallel \alpha}} \nonumber \\
   & -  \sum_\alpha  \fs{\nabla \ln B \cdot \bm Q_{ \alpha}^\perp } =  \fs{\mathbb{S}_{\epsilon}},
\end{eqnarray}
with $\bm Q = \sum_\alpha \bm Q_\alpha$ the total energy flux and $ \mathbb{S}_{\epsilon} = \sum_\alpha \mathbb{S}_{\epsilon \alpha}$. Here, the contribution from the collisional heating ($Q_{\alpha \beta} $) vanishes due to the energy conservation of the collision operator \cite{frei2025vspec}. We remark that the second and third terms on the left hand-side in \eqref{eq:steadystateheat} are smaller by at least a factor $\rho_s / L_{\parallel}$ ($L_{\parallel} \sim q R$ being the typical connection length along the magnetic field) compared to the first one, which is dominated by the turbulent $\bm E \times \bm B$ heat flux (see \sref{sec:sec8}). Hence, the two former terms can be neglected relative to the latter in \eqref{eq:steadystateheat}. Thus, the thermal energy balance equation reduces, at QSS, to

\begin{equation} \label{eq:steadystateheat2}
   \frac{1}{V^\prime} \frac{d}{d \psi} \left(V' \fs{\bm Q \cdot \nabla \psi }  \right)   \simeq  \fs{\mathbb{S}_{\epsilon}},
\end{equation}
omitting the $\avgt{\cdot}$ notation for simplicity. Integrating from the edge ($\rho_{\mathrm{pol}} \simeq 0.9$) and to the separatrix ($\rho_{\mathrm{pol}} \simeq 0.99$), \eqref{eq:steadystateheat2} implies that that the time-averaged total energy flux, $\fs{\bm{Q}  \cdot \nabla \psi}$, is constant across the edge, if $\mathbb{S}_{\epsilon } = 0$. Otherwise, it is equal to $3 \sum_\alpha S_{n_\alpha} T_{S \alpha} / 2$.

We verify qualitatively the particle and energy balance equations, \eqref{eq:densitybalanceintegrated} and \eqref{eq:steadystateheat2} respectively, in our simulations. \fref{fig:fig3} shows the radial profiles of the surface-integrated and time-averaged total electron particle flux, $\Gamma_e = V^\prime \fs{\bm{\Gamma}_\alpha \cdot \nabla \psi} $, and of the total heat flux, $P_{\mathrm{tot}} = V^\prime \fs{ \bm{Q} \cdot \nabla \psi}$, for the different time slices (see \tref{table:simsummary}) at QSS. In the absence of a localized density source, the radial fluxes remain approximately constant, consistent with \eqref{eq:densitybalanceintegrated} and \eqref{eq:steadystateheat2}, and increase as the L-H transition is approached due to the rise of the inner edge boundary. When the localized density source is introduced (indicated by the shaded region) for $t = 4.8~\mathrm{s}$ only, the particle flux $\Gamma_e$ is strongly reduced in the edge and subsequently increases by approximately $S_{n_\alpha} = \int dV \, \mathbb{S}_{n_\alpha} \approx 1.6 \times 10^{22}~\mathrm{s}^{-1}$ from $\rho_{\mathrm{pol}} \gtrsim 0.99$ towards the separatrix. In contrast, the total heat flux $Q_{\mathrm{tot}}$ remains nearly constant across the edge in all cases, since $\mathbb{S}_{\epsilon} = 0$.

\fref{fig:fig3} demonstrates that the present simulations are qualitatively at QSS with respect to particle and energy transport in the edge. Note that in practice, accurately satisfying these balance equations is limited by numerical and statistical noise, as the relevant quantities are computed in post-processing. In the following sections, statistical analyses are performed over a time window of $0.1~\mathrm{ms}$ at QSS.

\section{Outboard Midplane Profile and Radial Electric Field Validation}
\label{sec:validation}

In this section, we analyze the outboard midplane (OMP) profiles and validate them against experimental measurements for the different time slices approaching the L-H transition. In particular, we validate the density ($n_e$), electron ($T_e$) and ion ($T_i$) temperatures OMP profiles in \sref{subsec:denstempvalidation}. Most importantly and due to its relevant in the L-H transition physics, we validate the OMP profiles of the radial electric field ($E_r$) in \sref{subsec:ervalidation}. The normalized OMP gradients are reported in \sref{sec:appendixd}.

\label{subsec:ervalidation}

\subsection{OMP Density and Temperature Profile Validation}
\label{subsec:denstempvalidation}

\begin{figure}[h]
    \centering
    \includegraphics[scale=0.55]{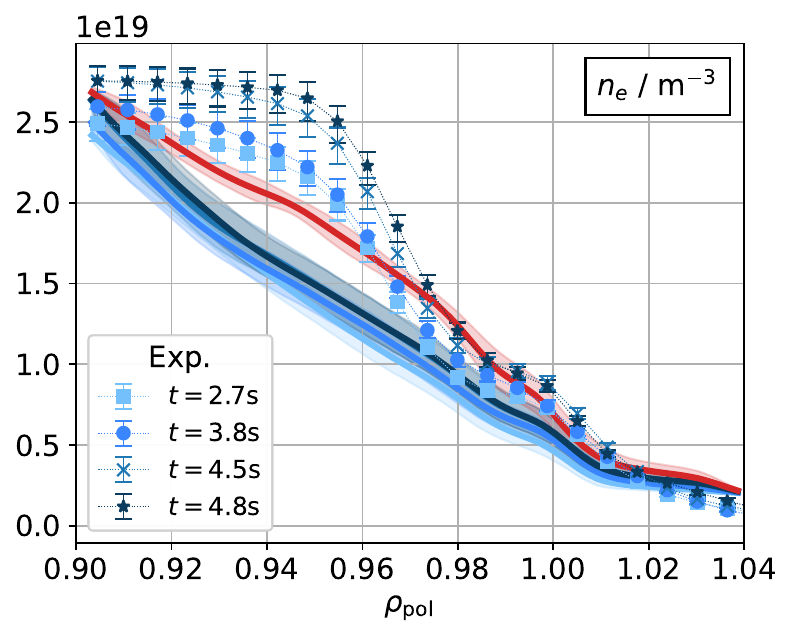}
    \caption{OMP density $n_e$ profiles plotted as a function of the normalized flux surface label $\rho_{\mathrm{pol}}$. The \texttt{GENE-X} results are shown by the colored lines (\tref{table:simsummary}) and the shaded colored indicate the standard deviations associated with the toroidal and time averages performed at QSS. Experimental measurements (from TS and ECE diagnostics) are also shown.}
    \label{fig:densomp}
\end{figure}

The OMP density profiles are obtained by interpolating the data at the OMP, followed by averaging over the toroidal direction and over $0.1~\mathrm{ms}$ at QSS. The resulting profiles are compared with experimental measurements in \fref{fig:densomp}. First, the agreement with experiments around $\rho_{\mathrm{pol}} = 0.9$ arises from the Dirichlet boundary condition (see \tref{table:simsummary}). Without the density source $S_n$, the simulated profiles systematically underestimate the experimental values, consistent with previous L-mode simulations \cite{michels2021, ulbl2023,frei2025}. They also exhibit only minor differences between the different time slices due to the slow density evolution approaching the L-H transition in this discharge. This systematic underestimation arises from the lack of edge density sources, such as neutral-particle ionization. Indeed, when $S_n$ is introduced at $t = 4.8~\mathrm{s}$, the edge density (and the associated gradients) near the separatrix increase, leading to an improved agreement with experiment, particularly in the range $0.96 \lesssim \rho_{\mathrm{pol}} \lesssim 1.02$. At the same time, the density gradients, shown in \fref{fig:ompgradients}, are reduced for $\rho_{\mathrm{pol}} \lesssim 0.96$, yielding a turbulence stabilization and a reduction of the radial particle flux (see \fref{fig:fig3}) producing the observed profile buildup for $\rho_{\mathrm{pol}} \lesssim 0.98$. Although the introduction of $S_n$ improves the agreement near the separatrix, the simulated density remains below the experimental values for $\rho_{\mathrm{pol}} \lesssim 0.96$. Nevertheless, the predicted $n_e$ profiles closely match the experimental measurements inside the separatrix, where the $E_r$ well is located.

\begin{figure}[h]
    \centering
    \includegraphics[scale=0.55]{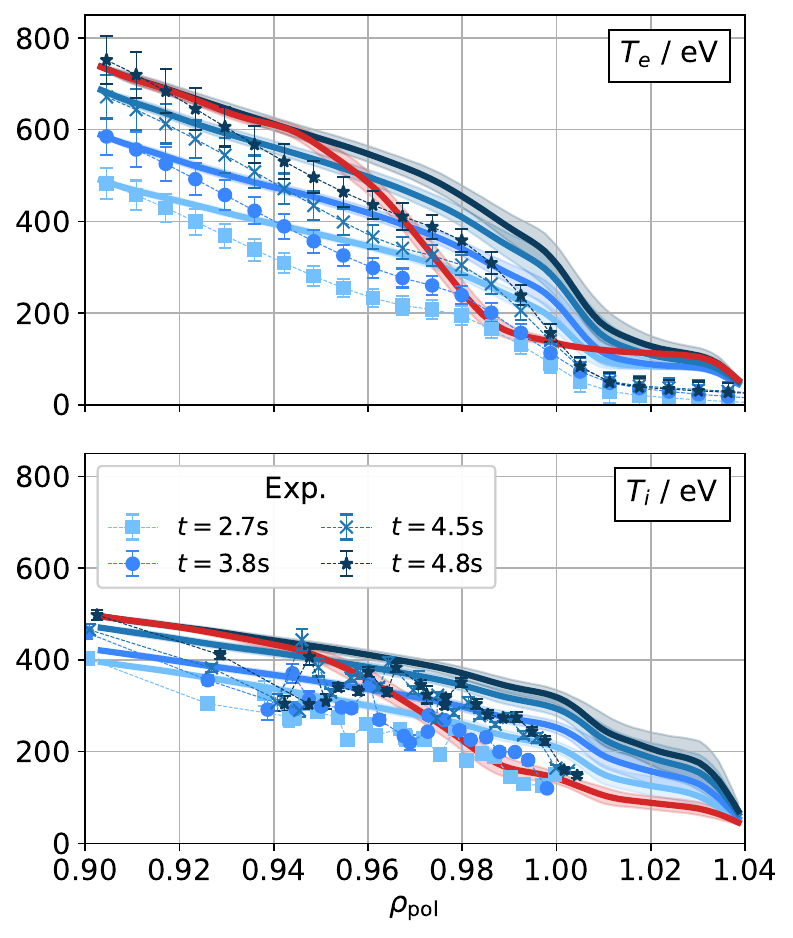}
    \caption{Same as \fref{fig:densomp} for the OMP electron $T_e$ (top) and ion $T_i$ (bottom) temperature profiles. Experimental measurements (from TS, ECE, and CX diagnostics) are also shown.}
    \label{fig:tempomp}
\end{figure}

\Fref{fig:tempomp} compares the OMP electron and ion temperature profiles, $T_e$ and $T_i$ respectively, with the experimental measurements. As observed previously, the agreement with experiments near the edge boundary primarily results from the imposed Dirichlet boundary conditions. The increase in input power approaching the L-H transition is modeled through a stepwise increase of the edge boundary temperature. In the absence of the density source $S_n$, both $T_e$ and $T_i$ are overestimated, with the discrepancy being more pronounced for $T_e$ in the edge region. The $T_i$ profiles nevertheless exhibit better agreement with the experimental data. The simulations correctly reproduce $T_e > T_i$ in the edge, consistent with ECRH-dominated heating, with $T_i \gtrsim T_e$ near the separatrix.
The $T_e$ and $T_i$ profiles shown in \fref{fig:tempomp} indicate that a similar instability drive is expected in cases without the density source $S_n$, since the normalized temperature gradients remain comparable (not shown).

The simulated SOL temperatures are also systematically higher than experiments. This overestimation can be attributed to several modelling assumptions such as the use of imposed Dirichlet boundary conditions, 
the absence of explicit temperature sinks in the SOL (the density source is located inside the separatrix), 
and the treatment of parallel heat transport between the outer midplane and the divertor. In particular, SOL parallel heat transport is highly sensitive to the sheath boundary conditions and the simplified divertor boundary conditions employed here (see \sref{subsec3.2}) are therefore likely to contribute 
to the large and radial shift (compared to experiments) of the SOL temperature gradients observed in the simulations.

When the density source $S_n$ is introduced at $t = 4.8\,\mathrm{s}$, it effectively acts as a localized temperature sink, reducing the separatrix temperatures close to the experimental values near the separatrix. However the corresponding gradients are also reduced such that the region of steepest gradients are slightly shifted inward compared to experiments. 

\subsection{OMP $E_r$ Profile Validation}
\label{subsec:ervalidation}

\begin{figure*}[h]
    \centering    
    \includegraphics[scale=0.55]{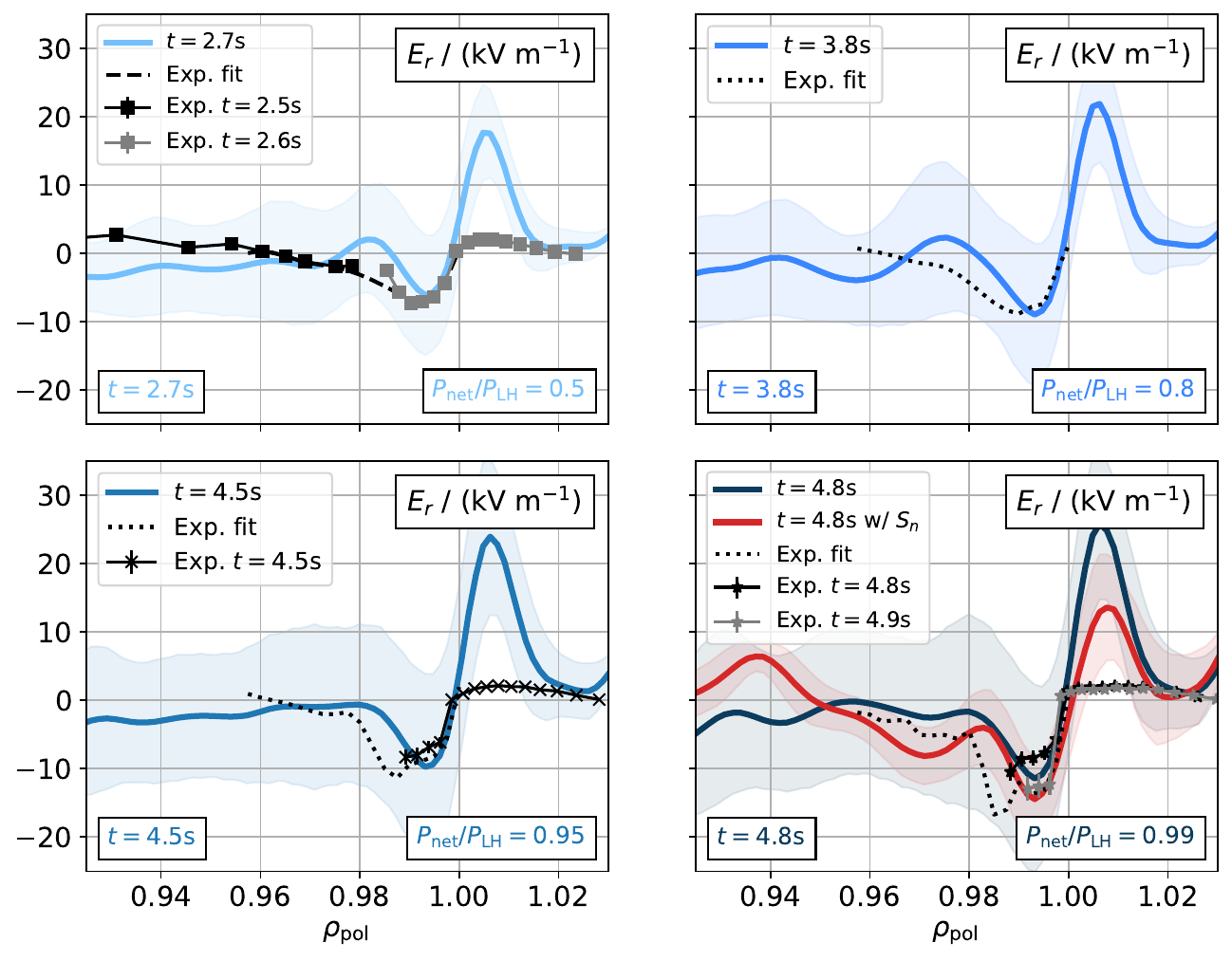}
    \caption{OMP $E_r$ profiles at different time slices approaching the L-H transition (from top left to bottom right). Solid colored lines show \texttt{GENE-X} simulation results (\ref{table:simsummary}). Experimental data include raw measurements at the corresponding time slices (\markerSquare[], \markerCircle[], \markerCross[], \markerStar[] markers) and time-averaged fitted profiles from CXRS and reflectometry (dotted black lines and taken from~\cite{bonanomi2024}). At $t=4.8$~s, the results without (\textcolor{color4p8}{\rule{2ex}{1.5ex}}) and with (\textcolor{tabred}{\rule{2ex}{1.5ex}}) $S_n$ are compared. Measurement of $E_r$ in H-mode at $t=4.9$~s is also shown for comparison. Colored shaded regions indicate the standard deviation associated with toroidal and time averages.}
    \label{fig:fig5}
\end{figure*}

Since the radial electric field $E_r$ is widely believed to play a central role in the L-H transition physics \cite{sauter2011,schmitz2012,plank2022}, it is essential to validate predictive models against experimental measurements in the vicinity of the transition. For the present $E_r$ validation, we use the same experimental dataset as in \cite{bonanomi2024} with the $E_r$ profiles, inferred from a combination of Doppler reflectometry \cite{conway2010} and charge-exchange (CX) \cite{plank2023} measurements, averaged over different time slices and employed in the local \texttt{GENE} analysis \cite{bonanomi2024} to impose the $\bm E \times \bm B$ shear.

In the simulations, $E_r$ is obtained from the negative radial gradient of the electrostatic potential, $E_r = - \nabla \psi \cdot \nabla \phi_1 / \left| \nabla \psi \right|$, and interpolated at the OMP. We remark that the simulated $E_r$ evolves self-consistently from the GK quasineutrality condition and includes contributions from both long wavelength neoclassical (NC) and short wavelength turbulent components \cite{frei2025}.

\Fref{fig:fig5} compares the $E_r$ profiles obtained from \texttt{GENE-X} with the experimental measurements for different time slices approaching the L-H transition. In all cases, the simulations exhibit a pronounced negative $E_r$ well near the separatrix ($\rho_{\mathrm{pol}} \simeq 0.99$). Inside the edge region ($\rho_{\mathrm{pol}} \lesssim 0.98$), $E_r$ remains nearly constant and negative, with values around $-1$~kV/m. Across the separatrix, $E_r$ increases sharply, changing sign (from negative to positive) near $\rho_{\mathrm{pol}} \simeq 1$ and reaching values of up to $20$~kV/m in the near SOL, before decreasing further outward (due to the boundary conditions).

We find excellent qualitative and quantitative agreement with experimental measurements around $\rho_{\mathrm{pol}} \sim 0.99$. More precisely, \texttt{GENE-X} accurately reproduces both the depth and width of the well as well as the radial location of the $E_r$ well with experiments despite the radial misalignment between the simulated and experimental profiles (see \fref{fig:tempomp}). Further inside the edge $\rho_{\mathrm{pol}} \lesssim 0.98$, experimental data are unavailable, preventing direct validation. The predicted $E_r$ profiles (and shear) are also consistent with those used in the local \texttt{GENE} analysis \cite{bonanomi2024}. In the SOL, however, the simulations systematically overestimate $E_r$: while measurements show peaks of order $\sim 3$~kV/m, \texttt{GENE-X} predicts values up to $20$~kV/m. This discrepancy is attributed to steep electron temperature gradients in the SOL~\cite{zholobenko2021b,frei2025}. Consistently, the presence of the density source $S_n$ at $t = 4.8$~s reduces the SOL $E_r$ to approximately $10$~kV/m by mitigating the radial $T_e$ gradient (see \fref{fig:ompgradients}).

The introduction of the density source at $t = 4.8$~s (corresponding to $P_{\mathrm{net}}/P_{\mathrm{LH}} = 0.99$) has several notable effects on $E_r$, as shown in the bottom-right panel of \fref{fig:fig5}. While the mean $E_r$ in the inner edge region ($\rho_{\mathrm{pol}} \lesssim 0.97$) remains qualitatively unchanged, the depth of the $E_r$ well increases. With $S_n$, $E_r$ reaches values of about $-15$~kV/m, comparable to experimental measurements in H-mode for the same discharge, as illustrated by the $E_r$ measurements at $t = 4.9$~s in H-mode. This demonstrates that the predicted depth of the $E_r$ well is consistent with experimental measurements close to the L-H threshold at AUG~\cite{sauter2011}.

The density source also has a strong impact on the amplitude of $E_r$ fluctuations, as inferred by the reduced standard deviation. In particular, the high-frequency fluctuations associated with eDW and TEM turbulence (see \sref{sec:sec7}) are substantially suppressed when $S_n$ is included, while low-frequency geodesic acoustic mode (GAM) oscillations persist (see \sref{subsec6.4}). As a consequence, the total $E_r$ (mean plus fluctuations) remains negative within the well, whereas in the absence of the source it can intermittently become positive. Finally, we note that both the depth and width of the $E_r$ well are sensitive to the properties of the density source as analyzed in \ref{sec:appendixb}.

\section{Radial Electric Field Analysis}
\label{sec:sec6}

Having validated the OMP $E_r$ profiles against experimental measurements, we now examine the composition of $E_r$ approaching the L-H transition. At AUG, $E_r$ is predominantly provided by the ion pressure gradient (diamagnetic term) at the edge in H-mode \cite{viezzer2013}, with the poloidal rotation at the neoclassical level. In contrast, toroidal and poloidal flows are non-negligible in the radial force balance under L-mode conditions \cite{plank2023}. A similar behavior is found here prior to the L-H transition: while $E_r$ is mainly sustained by the combination of main ion pressure gradient and toroidal rotation in the edge region ($\rho_{\mathrm{pol}} \lesssim 0.98$), the $E_r$ well, located near inside the separatrix $\rho_{\mathrm{pol}} \sim 0.99$, is governed by poloidal flows that strongly deviate from local NC predictions, indicating an important role of turbulence-driven flows.

We use the radial force balance introduced in \sref{subsec6.1} to analyze the different contributions to $E_r$. We compare the simulated poloidal flows with NC estimates in \sref{subsec6.2} and assess the evolution of the shear layers in \sref{subsec6.3}. Finally, the GAM activity near the L-H transition is discussed in \sref{subsec6.4}. This section is complemented by \ref{sec:appendixb} which examines the impact of the density source properties on $E_r$.

\subsection{Radial Force Balance Analysis}
\label{subsec6.1}

\begin{figure}[h]
    \centering
    \includegraphics[scale=0.55]{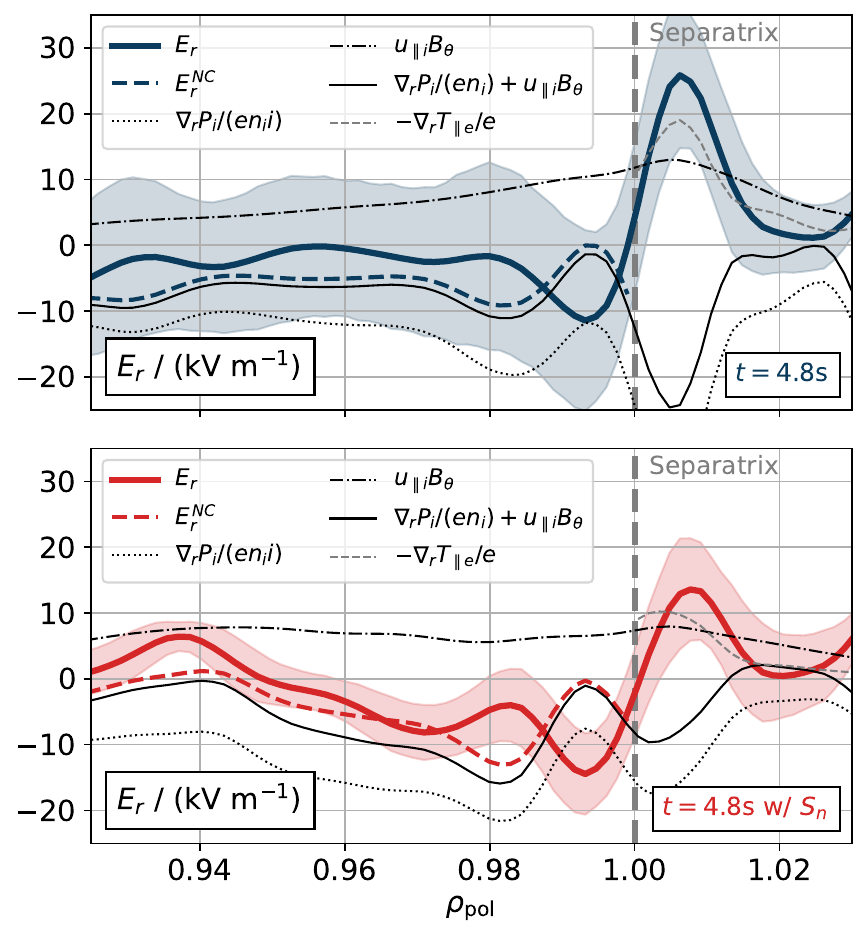}
    \caption{Decomposition of $E_r$ at $t=4.8$~s (or $P_{\mathrm{net}} / P_{\mathrm{LH}} = 0.99$) according to the radial force balance, \eqref{eq:forcebalance}. The \texttt{GENE-X} results without (top) and with (bottom) the density source $S_n$ are shown. The diamagnetic (black dotted line), toroidal flow (black dashed line), and their combined (black solid line) contributions are overlaid. The SOL estimation, $E_r \sim - \partial_r T_{\parallel e} / e $ \cite{frei2025}, is indicated by the gray dashed line, while the NC predictions, $E_r^{\mathrm{NC}}$, are shown by the dashed colored lines. The separatrix position is shown by the vertical dashed gray lines.}
    \label{fig:radialforcebalance}
\end{figure}

Although the radial force balance is intrinsically satisfied in full-$f$ formalism \texttt{GENE-X} \cite{frei2025}, it remains a useful diagnostic for disentangling the individual contributions to $E_r$. The radial force balance between $E_r$, the ion pressure gradient, and the (toroidal and poloidal) flows is given by \cite{frei2025}

\begin{equation} \label{eq:forcebalance}
E_r = \frac{1}{ q_i n_i} \frac{\partial P_i}{\partial r} -   U_{i \theta} B_\phi +  U_{i \phi} B_\theta, 
\end{equation} 
where $P_i$ is the total ion pressure (perpendicular and parallel pressures), $U_{i \theta}$ ($U_{i \phi}$) is the poloidal (toroidal) component of the ion flow $\bm U_{i}$ and $B_{\theta}$ ($B_{\phi}$) is the poloidal (toroidal) component of $\bm B$. In \eqref{eq:forcebalance}, it is assumed that the ion pressure tensor is isotropic \cite{frei2025}, an approximation well satisfied in the present simulations. From \eqref{eq:forcebalance}, we identify the first term as  the main ion radial pressure gradient (diamagnetic term) $\partial_r P_i / (q_i n_i)$, the second and third terms as the poloidal and toroidal components of the Lorentz force, $- U_{i \theta} B_\phi$ and $U_{i \phi} B_\theta$, respectively. The self-consistent ion flow $\bm U_{i}$ is given by the sum of the total gyrocenter drift and the classical magnetization term \cite{frei2025}, such that

\begin{equation} \label{eq:upol}
\bm U_{i  } = \frac{1}{n_i} \int d W \dot{\bm{R}} f_i   + \frac{c}{q_i n_i} \left(\nabla \times \bm m_i \right),
\end{equation}
where $\bm m_i =  - P_{\perp i} \bm b / B$ is the classical magnetization proportional to the perpendicular pressure \cite{hazeltine2013}. Note that $\bm U_{i}$ contains both NC and turbulence-driven contributions. From \eqref{eq:upol}, the poloidal and toroidal flows can be obtained by $U_{i \theta} = \bm b \times \nabla \psi  \cdot \bm U_i / \left| \nabla \psi \right|$ and $U_{i \phi} \simeq \bm b \cdot \bm U_i$, respectively.
To assess the contribution turbulence-driven poloidal flow to $E_r$, we also calculated $E_r$ from the force balance assuming that the ion poloidal flow is at NC level, i.e. $ U_{i \theta}\simeq U_{i \theta}^{\mathrm{NC}} = K_1 \partial_r T_i / (q_i B)$ \cite{kim1991,plank2022}, such that

\begin{equation} \label{eq:ernc}
E_r^{\mathrm{NC}} = \frac{T_i}{q_i} \left[ \frac{\partial}{\partial r} \ln n_i + (1 - K_1) \frac{\partial}{\partial r} \ln T_i \right]  +  U_{i \phi} B_\theta.
\end{equation}
Here, $ K_1$ is a coefficient that strongly depends on the inverse aspect ratio and the normalized ion collisionality $\nu_i^*$ \cite{kim1991}. The profile of $\nu_i^*$ at $t=4.8$~s (not shown) reveals that $\nu_i^* \simeq 0.5$ ($K_1 \simeq 0.4$) is in the banana regime \cite{plank2023} without a density source, but transitions to the plateau regime when $S_n$ is included where $1 \lesssim \nu_i^* \lesssim 3$ due to the lower edge temperature in this case. However, $K_1$ changes sign (from positive to negative) when $\nu_i^* \simeq 4.5$. Thus, $K_1 > 0$ in all the present simulations. 

We focus the present analysis near the L-H transition at $t=4.8$~s ($P_{\mathrm{net}} / P_{\mathrm{LH}} = 0.99$) and decompose $E_r$ according to \eqref{eq:forcebalance}. \Fref{fig:radialforcebalance} shows the results with and without the density source $S_n$ at the OMP position in the edge and SOL regions. In all cases, the diamagnetic approximation, $E_r \simeq \partial_r P_i /(q_i n_i)$, fails to reproduce the simulated $E_r$ well, highlighting the non-negligible role of toroidal and poloidal flows in $E_r$ approaching the L-H transition. Accounting for the toroidal flow contribution primarily shifts the diamagnetic estimate upward improving the agreement with $E_r$ only in the edge region ($\rho_\mathrm{pol} \lesssim 0.98$), but not near the separatrix. Similarly, the NC approximation (computed using the \texttt{GENE-X} profiles) provides a reasonable description of $E_r$ inside the edge in both cases. However, both approximations break down near the $E_r$ well: a systematic radial offset between the extrema of the diamagnetic term and the minimum of the $E_r$ well ($E_{r, \mathrm{min}}$) is noticeable. 

By construction, the difference between $\partial_r P_i / (q_i n_i) + U_{i\phi} B_\theta$ and $E_r$ corresponds to the poloidal flow contribution $-U_{i\theta} B_\phi$ in \eqref{eq:forcebalance}. Therefore, \fref{fig:radialforcebalance} demonstrates that the $E_r$ well structure is determined by the poloidal flows as the diamagnetic and toroidal flow contributions nearly cancel out at the location of $E_{r, \mathrm{min}}$. The same observation holds for the other time slices, but not shown. Remarkably, this balance between diamagnetic and toroidal terms in the $E_r$ well persists in the presence of $S_n$, indicating that the balance between these two contributions to $E_r$ is unaffected by the change of the diamagnetic term induced by $S_n$. This balance is also confirmed when the parameters (amplitude and position) of $S_n$ are varied as demonstrated in \sref{sec:appendixb}. Therefore, the deeper $E_r$ well observed with $S_n$ is due to the strengthening of the poloidal flows. Finally, we note that $E_r \sim - \partial_r T_{\parallel e} / e$ in the SOL due to the parallel electron dynamics \cite{frei2025}. 

\subsection{Edge Poloidal Flows}
\label{subsec6.2}

\begin{figure}[h]
    \centering   
    \includegraphics[scale=0.55]{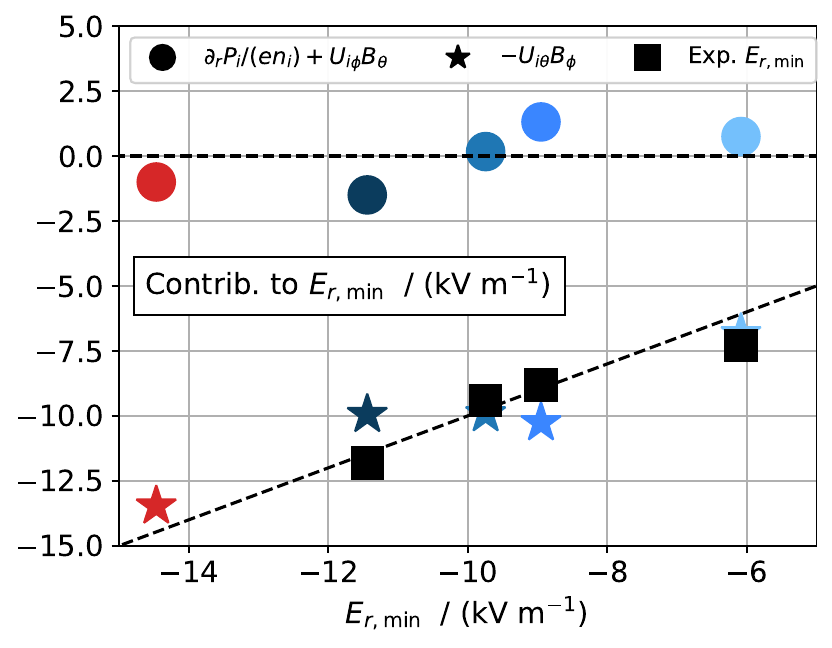}
    \caption{Individual contributions to $E_{r, \mathrm{min}}$ in the radial force balance equation, \ref{eq:forcebalance}, and plotted as a function of $E_{r, \mathrm{min}}$: $\partial_r P_i / (q_i n_i) + U_{i \phi} B_\theta$ (\markerCircle[]), and $-U_{i \theta} B_\phi$ (\markerStar[]). For comparison, the experimental $E_{r, \mathrm{min}}$, calculated from \fref{fig:fig5}, are also shown (\markerSquare[]).  $P_{\mathrm{net}} / P_{\mathrm{LH}}$ increases from right to left (see \tref{table:simsummary}).}
    \label{fig:ermin}
\end{figure}

The importance of poloidal flows in setting the depth of the $E_r$ well can be better visualized in \fref{fig:ermin} where the different contributions to $E_{r, \mathrm{min}}$ are plotted individually as a function of $E_{r, \mathrm{min}}$ itself for the different times slices. Experimental values of $E_{r, \mathrm{min}}$, obtained from \fref{fig:fig5}, are also shown for comparison. First, a remarkable agreement with the experimental $E_{r,\mathrm{min}}$ and the one predicted by \texttt{GENE-X} is observed approaching the L-H transition. However, $E_{r, \mathrm{min}}$ is found larger (in amplitude) with the density source despite better agreements of the $n_e$ profile with experiments. Second, the contribution of $\partial_r P_i / (q_i n_i) + U_{i \phi} B_\theta $ in $E_{r, \mathrm{min}}$ remain close to $0$~kV m$^{-1}$ and negligible compared to $- U_{i \theta} B_\phi$. As a consequence, $E_{r, \mathrm{min}}$ shows a strong correlation with the poloidal flow contribution, $-U_{i \theta} B_\phi$. This is particularly visible between $t= 2.7$~s and $t= 4.8$~s, with and without the density source, where $E_{r, \mathrm{min}}$ becomes more negative by approximately $4$~kV m$^{-1}$ between each case. We remark that the progressive deepening of the $E_r$ well between $t =3.8$~s, $4.5$~s and $4.8$~s can be attributed to the small decrease of $\partial_r P_i / (q_i n_i) + U_{i \phi} B_\theta $, while the poloidal flow (although the largest contribution) remains nearly constant between these time slices. In all cases, the diamagnetic term is of the same order as the radial component of the Lorentz force, $- U_{i \theta} B_\phi + U_{i \phi} B_\theta$ in \eqref{eq:forcebalance}, consistent with L-mode experimental observations \cite{plank2023}.

\begin{figure}[h]
    \centering   
    \includegraphics[scale=0.55]{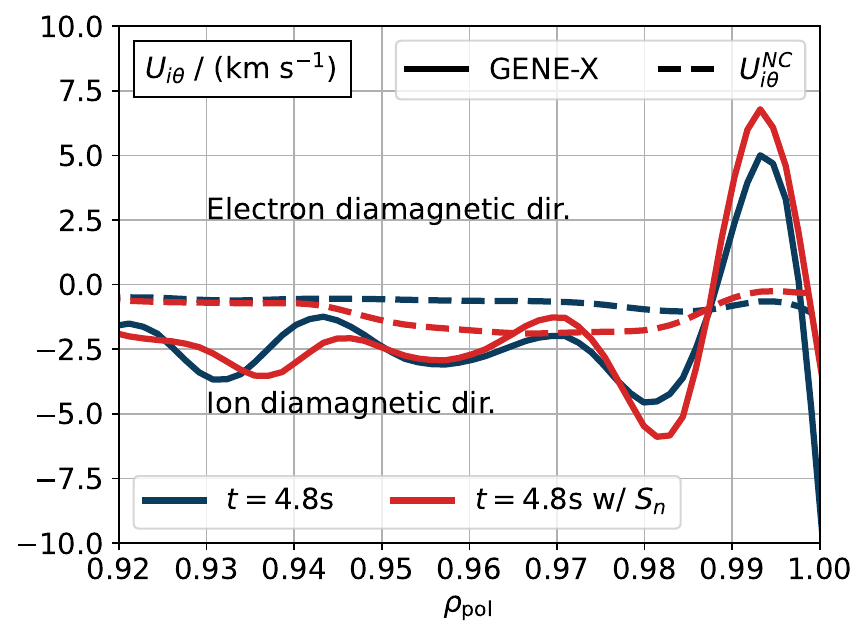}
    \caption{Poloidal flows $U_{i \theta}$ self-consistently calculated using \eqref{eq:upol} (solid lines) at the OMP obtained at $t=4.8$~s with (\textcolor{tabred}{\rule{2ex}{1.5ex}}) and without (\textcolor{color4p8}{\rule{2ex}{1.5ex}}) $S_n$. The NC predictions, $U_{i \theta}^{NC}$ \cite{kim1991} and calculated using \texttt{GENE-X} profiles, are also shown for comparison. Positive (negative) velocity indicates the electron (ion) diamagnetic direction.}
    \label{fig:poloidalflow}
\end{figure}

The mechanism driving the edge poloidal flows in the $E_r$ well shown in \fref{fig:radialforcebalance} can be inferred from \fref{fig:poloidalflow} where the OMP profiles of $U_{i\theta}$ (self-consistently calculated using \eqref{eq:upol}) and of the NC predictions $U_{i \theta}^{NC}$ (calculated using the \texttt{GENE-X} profiles) are compared at $t=4.8$~s, prior to the L-H transition. As observed, $U_{i \theta}$ is directed in the ion diamagnetic (negative) direction and has an amplitude of the order of $\sim -2$~km s$^{-1}$, which is consistent with $U_{i \theta}^{NC}$. However, towards the separatrix, $U_{i\theta}$ changes sign, from negative to positive (electron diamagnetic direction), and increases to $\sim 7$~km s$^{-1}$, while $U_{i \theta}^{NC}$ remains small and in the ion diamagnetic direction. 

 This indicates that $U_{i\theta}$ originates from non-NC effects. Indeed, nonlinear turbulence-driven mean (zonal) poloidal flows are known to play a central role in the onset of the L-H transition. These poloidal flows are driven by the radial gradient of the velocity stress (equivalent to the Reynold stress in the local limit), which is proportional to the gradient of the radial particle flux. In particular, the particle flux (and its gradient) is amplified around the $E_r$ well due to the presence of $S_n$ (see \fref{fig:fig3}). This nonlinear mechanism explains the larger mean (zonal) poloidal flow observed in \fref{fig:radialforcebalance} when the density source is included, hence deeper $E_r$ well. A more detailed analysis of the nonlinear interactions between turbulence and mean flows will be presented in a future publication. Finally, we note that the NC predictions used here are based on \textit{local} theory and therefore do not capture global effects \cite{landreman2014}, which can be significant in the plasma edge. For instance, the $E_r$ profile can modify the NC poloidal ion flow $U_{i\theta}^{\mathrm{NC}}$ and may even lead to a reversal of its sign if the $E_r$ shear is sufficiently strong \cite{kolesnikov2010}. A more comprehensive comparison with nonlocal and more accurate NC models is deferred to future work.

\subsection{Inner and Outer Shear Layers}
\label{subsec6.3}

\begin{figure}[h]
   \centering   
    \includegraphics[scale=0.55]{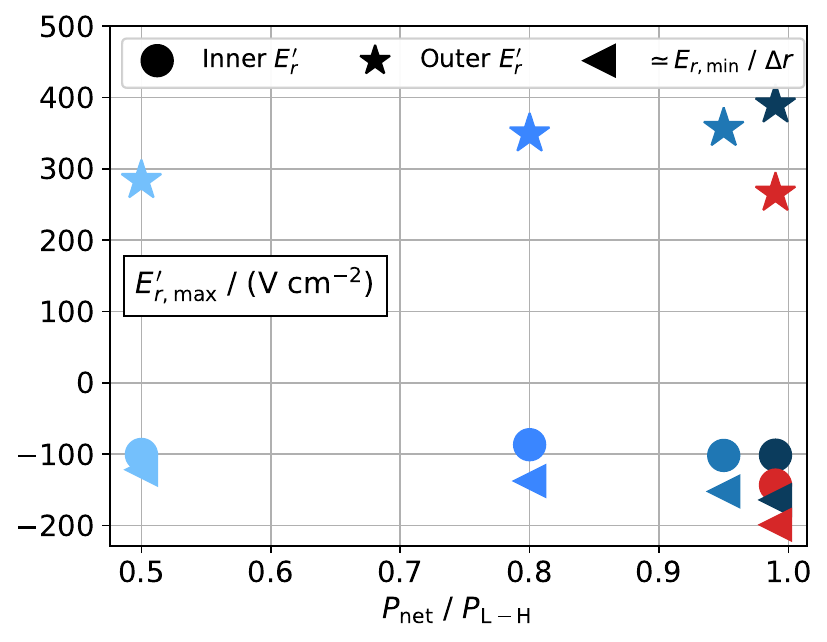}
\caption{Maximum of the inner (\markerCircle[black]) and outer (\markerStar[black]) $E_r$ shear layers around the $E_r$ well (displayed in \fref{fig:fig5}) obtained in the \texttt{GENE-X} simulations for different $P_{\mathrm{net}} / P_{\mathrm{LH}}$. For comparison, the inner shear linear estimates, $E_{r, \mathrm{min}} / \Delta r $ (\markerRightTriangle[black]), are also displayed for comparison.}
    \label{fig:ershearlayers}
\end{figure}

The existence of (inner and outer) shear layers around the $E_r$ well is of particular importance as these shear layers are considered central to the onset of the L–H transition \cite{cavedon2024}. \Fref{fig:ershearlayers} shows the evolution of the maximum $E_r$ shear, $E_r^\prime = dE_r/dr$, for both layers obtained from the \texttt{GENE-X} simulations. It is observed that the maximum of the inner shear remains approximately constant, while the outer shear increases when approaching the L–H transition. The latter is sensitive to the SOL $E_r$ and may, therefore, be overestimated in the present simulations. 

Because direct measurements of $E_r$ shear are challenging in experiments, a linear linear estimate, $E_r^\prime \simeq E_{r,\mathrm{min}}/\Delta r$ (with $\Delta r$ the full width at half maximum of the $E_r$ well), is typically used as a proxy for the inner shear \cite{sauter2011,viezzer2014,cavedon2024}. We compare this estimates with the \texttt{GENE-X} results in \fref{fig:ershearlayers}. Although the linear estimate tends to over predict the inner shear amplitude, it provides a reasonable qualitative approximation. Note that the simulated $\Delta r$ increases from about $5$ to $7$~mm approaching the L-H transition. The presence of the density source enhances the inner shear (deeper $E_r$ well and constant $\Delta r$) while reducing the outer shear due to a lower SOL $E_r$. While no direct validation is performed here, we note that typical values for the inner shear are of the order of $-75$~V / cm$^2$ in L-mode at AUG, consistent with \fref{fig:ershearlayers}, and are up to $-200$~V / cm$^2$ in H-mode \cite{schirmer2006}. 

Finally, the relative roles of inner and outer shear layers in turbulence mitigation (or suppression) remain inconclusive due to SOL conditions in the present simulations and more detailed analysis are needed to address open questions about the role of shear layers close to the L-H transition \cite{cavedon2024}. 

\subsection{Edge GAM Activity}
\label{subsec6.4}

Although their role in triggering the L–H transition remains unclear, 
GAMs which are finite-frequency oscillations of mean (zonal) flows can regulate and mitigate turbulent transport via flow shear if their amplitude is large enough. GAMs have characteristic frequencies of the order $f_{\mathrm{GAM}} \sim c_s/R$ and are typically observed in L-mode plasmas, with intensity increasing prior to the L–H transition in AUG \cite{conway2008,conway2011}, while being strongly reduced or suppressed in H-mode. In contrast to the core region, GAMs are weakly damped in the edge region \cite{sugama2006} and turbulence can provide their dominant drive via nonlinear interactions \cite{conway2008}, overcoming collisional and Landau damping mechanisms \cite{conway2021,frei2023}.

\begin{figure}[h]
   \centering   
    \includegraphics[scale=0.5]{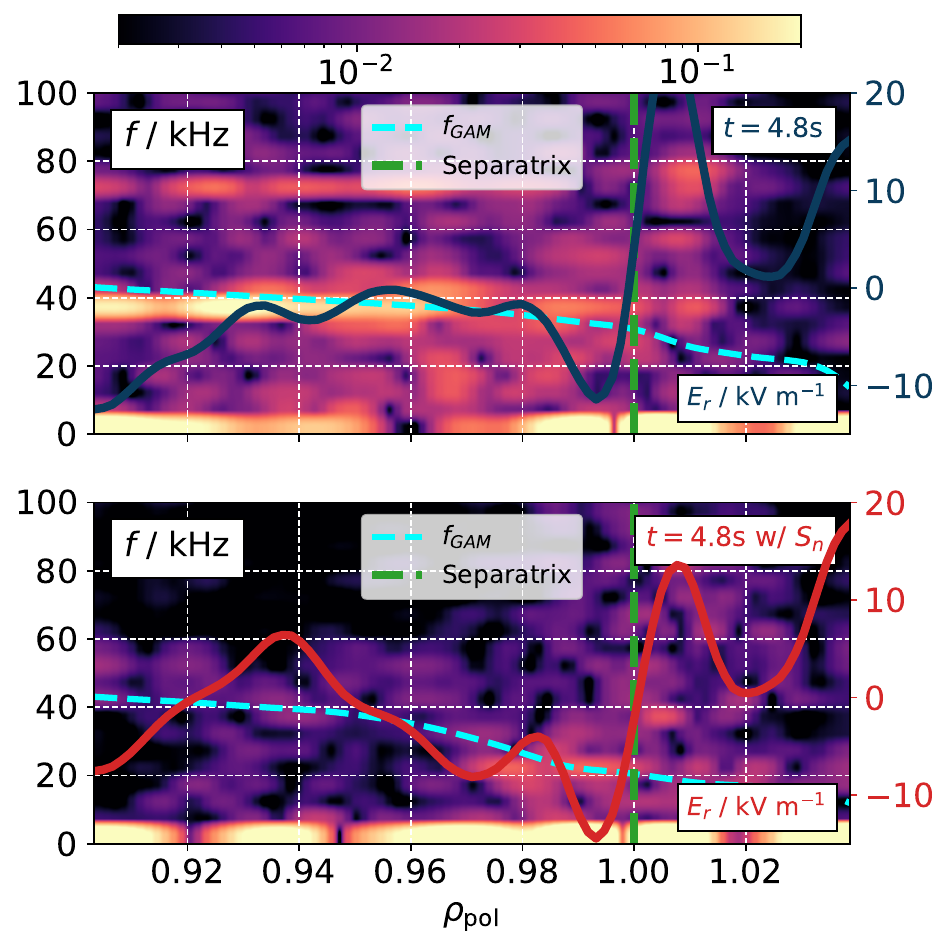}
    \caption{Radially resolved frequency power spectrum of $E_r$ (in kHz) at $P_{\mathrm{net}}/P_{\mathrm{LH}} = 0.99$, shown without (top) and with (bottom) density source. The linear estimate of the GAM frequency $f_{\mathrm{GAM}}$ \cite{sugama2006} is indicated by dotted red lines, and the separatrix position by green dashed lines.}
    \label{fig:fouriergam}
\end{figure}

In the present simulations, GAMs are identified as low-frequency ($f \sim 50$~kHz) temporal oscillations of the radial electric field $E_r$. The radial dependence of the $E_r$ frequency power spectrum, obtained with and without the density source, is shown in \fref{fig:fouriergam} at $t = 4.8$~s. For comparison, the analytical estimate of the GAM frequency $f_{\mathrm{GAM}}$ \cite{sugama2006} is also included. In the absence of $S_n$, the power spectrum exhibits a broadband structure at all radii, with a radially uniform dominant peak at $f \sim 40$~kHz, in close agreement with $f_{\mathrm{GAM}}$, although the theoretical $\sqrt{T_e}$ scaling of $f_{\mathrm{GAM}}$ is not recovered. Note that the simulated GAM frequencies are qualitatively consistent with measurements ($f_{\mathrm{GAM}} \sim 25$~kHz) at AUG \cite{conway2010,palermo2017}, but are larger due to higher temperatures in \texttt{GENE-X}. Radially localized zero-frequency bands are also observed in \fref{fig:fouriergam} and associated with zero-frequency (stationary) zonal flows. 

When $S_n$ is included, the spectrum becomes narrower and weaker, particularly in the edge region, consistent with reduced turbulence activity associated with the weaker density gradient (see \sref{sec:sec7}). The dominant spectral peak remains consistent with $f_{\mathrm{GAM}}$, but the GAM activity becomes radially localized around the $E_r$ well around $\rho_{\mathrm{pol}} \sim 0.99$. In addition, these GAMs oscillations originate near the location of the minimum of $E_r$ and propagate radially inward \cite{conway2011,palermo2017}, with a radial extend over the width of the $E_r$ well ($\sim 5$~mm). 

The GAM amplitude, which can be quantified by the standard deviation of $E_r$ (see \fref{fig:fig5})), remains smaller than the mean $E_r$ with $S_n$, but becomes comparable to it in the absence of the source. Although GAMs oscillations can affect the $E_r$ shearing rate, no clear modulation of turbulence amplitudes at $f_{\mathrm{GAM}}$ indicates a weak impact on turbulent transport.

\section{Turbulence Characterization}
\label{sec:sec7}

We characterize edge turbulence at $P_{\mathrm{net}} / P_{\mathrm{LH}} = 0.99$, close to the L--H transition. Previous linear flux-tube \texttt{GENE} analyses \cite{bonanomi2019,bonanomi2021,bonanomi2024} have shown that edge turbulence is predominantly driven by electron drift waves (eDWs), while trapped-electron modes (TEMs) remain subdominant, highlighting the importance of the parallel electron dynamics. A similar interplay between eDWs and TEMs is observed in the present simulations.

To investigate the turbulence characteristics, we compute the dispersion relation $\omega(k_y)$, which relates the real mode frequency $\omega$ to the binormal wavenumber $k_y$. The dispersion relation is obtained via a temporal Fourier analysis of the Fourier components of the electrostatic potential $\phi_1$. The Fourier amplitude is defined as $\hat{\phi}_1(k_y) =  \int_0^L \mathrm{d}y  \phi_1(y)\mathrm{e}^{-\mathrm{i} k_y y} / r$ and is evaluated on a given flux surface. Here, $y = r \theta$ is the poloidal arc length, $r = L / (2\pi)$ is the effective radius, $\theta$ is the geometrical poloidal angle, and $k_y = m / r$ corresponds to the poloidal mode number $m$. The frequency spectrum is then obtained from $\hat{\phi}_1(k_y,\omega) = \int \mathrm{d}t \, \hat{\phi}_1(k_y)\, \mathrm{e}^{\mathrm{i}\omega t}$. In our convention, positive frequencies ($\omega > 0$) correspond to modes propagating in the electron diamagnetic direction $\bm{k}_\perp \cdot \bm{V}_e^* > 0$, where $\bm{V}_e^* = - c \bm{B} \times \nabla_\perp P_e / (e n_e B^2)$ is the electron diamagnetic drift velocity. The dispersion relation, $\omega(k_y)$, is computed in the plasma co-moving frame rotating with the $\bm{E} \times \bm{B}$ drift averaged over a given flux surface \cite{ulbl2023}. To identify the dominant instabilities, linear local estimates of the real frequencies of eDWs and collisionless TEMs (cTEMs), $\omega_e^*$ and $\omega_{\mathrm{cTEM}}$ respectively \cite{frei2025}, are compared with $\hat{\phi}_1(k_y,\omega)$. Although derived from local linear theory, these estimates provide useful information for interpreting the turbulence dynamics.

\begin{figure}[h]
    \centering
    \includegraphics[scale=0.55]{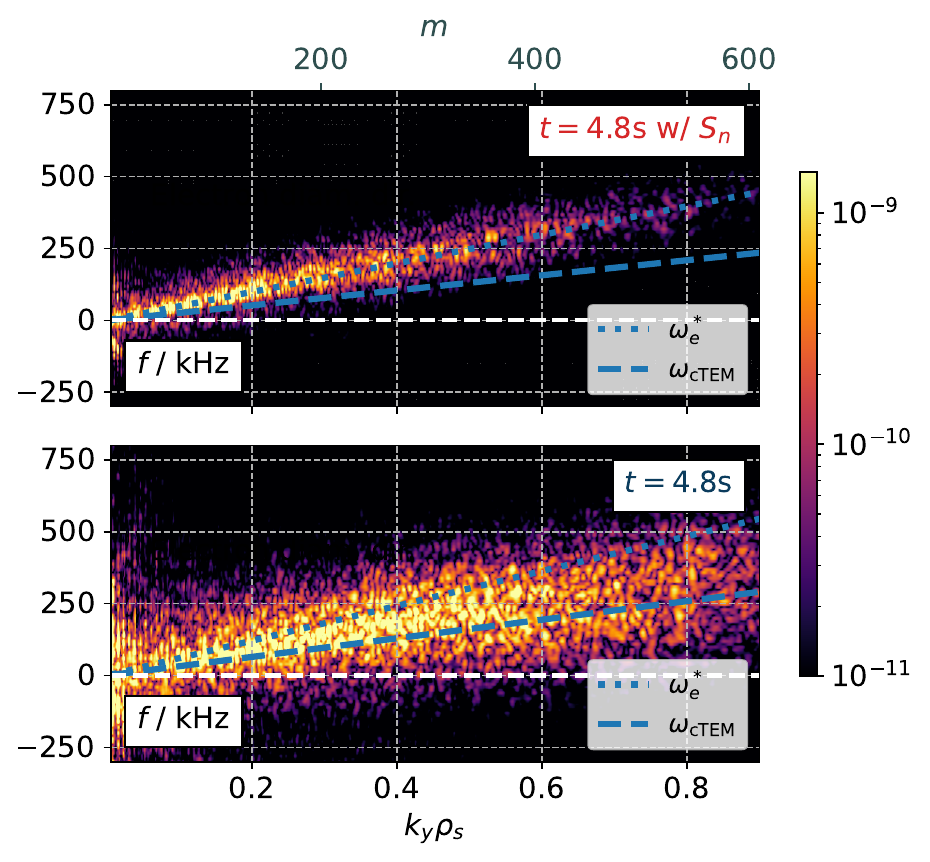}
    \caption{Dispersion relations $\omega(k_y)$ at $P_{\mathrm{net}} / P_{\mathrm{LH}} = 0.99$, shown with (top) and without (bottom) a density source at $\rho_{\mathrm{pol}} = 0.994$ inside the $E_r$ well. Here, $f = \omega / (2\pi)$ and the binormal wavenumber $k_y$ is normalized to the averaged sound Larmor radius $\rho_s$ on the flux-surface. The spectra are computed over a $0.2$~ms time window and averaged over the toroidal direction. The linear frequencies of the eDW and CTEM, $\omega_e^*$ and $\omega_{\mathrm{cTEM}}$, are overlaid as dotted and dashed blue lines, respectively. Positive frequencies correspond to the electron diamagnetic direction ($\bm{k}_\perp \cdot \bm{V}_e^* > 0$).}
    \label{fig:disprel}
\end{figure}

The dispersion relations obtained at $t = 4.8$~s ($P_{\mathrm{net}} / P_{\mathrm{LH}} = 0.99$) are shown in \fref{fig:disprel} with and without $S_n$. The analysis is performed on a flux surface within the $E_r$ well ($\rho_{\mathrm{pol}} = 0.994$). In both cases, turbulence propagates in the electron diamagnetic direction, with frequencies of order $150$~kHz and increasing with $k_y$. In the absence of $S_n$,  $\omega(k_y)$ is broadband, with a mean frequency lying between $\omega_e^*$ and $\omega_{\mathrm{cTEM}}$ and peaking around $k_y \rho_s \sim 0.4$ ($\rho_s$ is the ion sound Larmor radius averaged over the considered flux surface). When $S_n$ is included, $\omega(k_y)$ become significantly narrower and align more closely with $\omega_e^*$. Turbulence at $k_y \rho_s \gtrsim 0.4$ is strongly stabilized, and the peak shifts towards lower wavenumbers, around $k_y \rho_s \sim 0.2$. The alignment of the peaks of $\omega(k_y)$ with $\omega_e^*$ is consistent with linear flux-tube \texttt{GENE} analyses of the experimental profiles \cite{bonanomi2024}, which identify collisional eDWs as the dominant instability. Indeed, $\omega_{\mathrm{cTEM}}$ lies systematically below the peaks of $\omega(k_y)$, indicating that TEMs remain subdominant. Here, the collisionality is found to be in the banana and Plateau regimes and is, therefore, not large enough to fully suppress TEM turbulence. Similar results are obtained for the other time slices, although with lower amplitudes. This suggests that the fundamental nature of edge turbulence, dominated by eDWs, remains unchanged approaching the L-H transition, while its intensity increases. A more detailed linear stability analysis based on the \texttt{GENE-X} profiles could be performed and compared directly with \cite{bonanomi2024}, but it is beyond the scope of the present work.

\section{Transport Approaching the L-H Transition}
\label{sec:sec8}

Finally, we investigate the radial particle and heat transport approaching the L–H transition. This analysis is of primary importance since the edge ion heat flux, $Q_i$, is believed to play a decisive role in accessing H-mode \cite{ryter2014}. We compare the \texttt{GENE-X} predictions with the experimental ion and electron heat fluxes inferred from interpretative \texttt{ASTRA} simulations \cite{pereverzev2002,fable2013}, as well as with nonlinear flux-tube \texttt{GENE} results previously reported in \cite{bonanomi2024} based on the experimental profiles.

The total (per species) radial \textit{gyrocenter} heat flux, $Q_\alpha = \bm{Q_\alpha} \cdot \nabla \psi$, defined in \sref{sec:sec3}, can be decomposed into different contributions according to $\dot{\bm R} \cdot \nabla \psi \simeq \bm E_1 \times \bm B / B^2  \cdot \nabla \psi +  \bm V_{D \alpha}\cdot \nabla \psi +  v_\parallel \nabla A_{\parallel 1} \times \bm b / B  \cdot \nabla \psi $. Here, we identify the different perpendicular drifts as the electrostatic (es) $\bm E \times \bm B$ drift (neglecting the second order correction to $\phi$), diamagnetic (or magnetic velocity-dependent) diamagnetic (diam) drift $ \bm V_{D \alpha}$, and flutter (em) drift $v_\parallel \nabla A_{\parallel 1} \times \bm B / B^2$, respectively. This enable us to write $Q_\alpha = Q_{\alpha}^{\mathrm{es}} + Q_\alpha^{\mathrm{diam}} + Q_\alpha^{\mathrm{em}}$. From $Q_\alpha$, the total power, $P_\alpha$, crossing a given flux surface is given by $P_\alpha = V^\prime \fs{Q_\alpha} $. $P_\alpha$ follows a similar decomposition as $Q_\alpha$. e.g., $P_\alpha^{\mathrm{es}} = V^\prime \fs{Q_\alpha^{\mathrm{es}}} $ (and similarly for $P_\alpha^{\mathrm{diam}}$ and $P_\alpha^{\mathrm{em}}$).   

\begin{table}[t]
\centering
\begin{tabular}{ccc}
 $P_{\mathrm{net}} / P_{\mathrm{LH}} = 0.99$ & $P_e$ / MW &  $P_i$ / MW \\
\hline
\hline
\texttt{GENE} & $2.0$ & $1.2$   \\
\texttt{GENE-X} wo/ $S_n$ & $3.28$ & $1.46$   \\
\texttt{GENE-X} w/ $S_n$ &  $1.01$ & $0.73$   \\
Exp. (\texttt{ASTRA}) & $1.64$  &  $0.88$ \\
\hline
\end{tabular}
\caption{Total electron ($P_e$) and ion ($P_i$) powers at $P_{\mathrm{net}} / P_{\mathrm{LH}} = 0.99$ ($t = 4.8$~s) crossing the $\rho_{\mathrm{pol}} = 0.98$ flux surface obtained from nonlinear flux-tube \texttt{GENE}, \texttt{GENE-X} (without and with $S_n$), and experiments from interpretative \texttt{ASTRA}. \texttt{GENE} and \texttt{ASTRA} data are taken from \cite{bonanomi2024}.}
\label{table:powers}
\end{table}

\begin{figure}[h]
    \centering
    \includegraphics[scale=0.55]{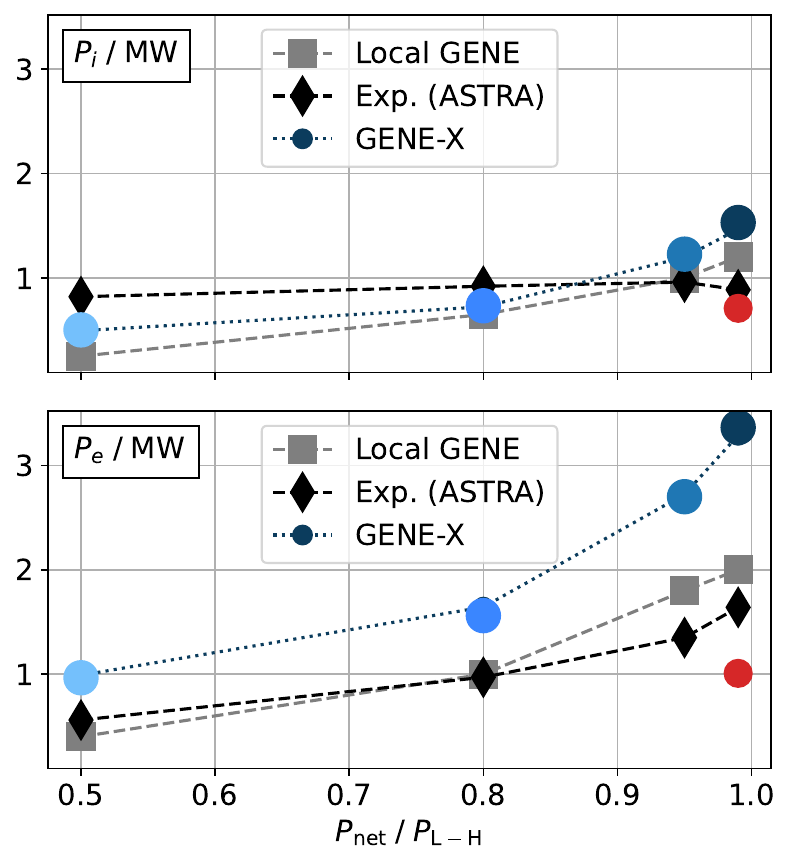}
    \caption{Evolution of the \textit{total} ion and electron powers, $P_i$ (top) and $P_e$ (bottom) respectively, approaching the L-H transition obtained from \texttt{GENE-X} (dotted lines with colored \markerCircle[black]), nonlinear local flux-tube \texttt{GENE} (dashed line with \markerSquare[gray], data from \cite{bonanomi2024}), and experiments inferred from \texttt{ASTRA} (dashed line with \markerDiamond[black], data from \cite{bonanomi2024}). The $\rho_{\mathrm{pol}} = 0.98$ flux surface (same as in \cite{bonanomi2024}) are used to calculate $P_i$ and $P_e$. See \tref{table:simsummary} for the color code.} 
    \label{fig:fluxesedges}
\end{figure}

\Fref{fig:fluxesedges} shows the evolution of the ion and electron powers, $P_i$ and $P_e$, across the $\rho_{\mathrm{pol}} = 0.98$ flux surface (same as in \cite{bonanomi2024}) approaching the L–H transition. The results from nonlinear flux-tube \texttt{GENE} simulations and experimental estimates inferred using interpretative \texttt{ASTRA} are also shown \cite{bonanomi2024}. As observed, the \texttt{GENE-X} simulations reproduce the characteristic increase of both $P_i$ and $P_e$ observed experimentally as the L–H transition is approached with $P_e > P_i$, reflecting the dominance of ECRH heating in this discharge (see \fref{fig:fig1}). The Fourier spectra of the convective and conductive ion and electron fluxes are shown in \sref{sec:appendixc}.

The increase in heat transport is in qualitative agreement with experiments and with flux-tube \texttt{GENE} simulations. However, in the absence of $S_n$, \texttt{GENE-X} systematically predicts larger heat fluxes, in particular for $P_e$. In contrast, flux-tube \texttt{GENE} shows a closer agreement with the experimental $P_e$. For the ion heat transport, $P_i$ is overestimated by \texttt{GENE-X} for $P_{\mathrm{net}} / P_{\mathrm{LH}} \gtrsim 0.9$, while improved agreement with experiments is obtained closer to the L-H transition. However, without density source, \texttt{GENE-X} yields an ion edge heat flux $Q_i$ ($1.46$~MW), which approximately $1.6$ larger than \texttt{ASTRA} prediction ($0.88$~MW). This difference is larger for $P_e$.

Overall, \texttt{GENE-X} predicts higher total heat transport for both ions and electrons compared to flux-tube simulations, with the discrepancy being more pronounced for $P_e$. We emphasize that, unlike the present full-$f$ simulations in which profiles evolve self-consistently, local flux-tube \texttt{GENE} calculations use experimental profiles as inputs and require an externally imposed $\bm E \times \bm B$ shearing rate inferred from measurements. Although these profiles are qualitatively similar to those predicted by \texttt{GENE-X} (see \sref{sec:validation}), differences in the normalized gradients and thus in the resulting turbulent fluxes may account for the discrepancies between \texttt{GENE} and \texttt{GENE-X}, particularly for $P_e$. The local parameters calculated from the \texttt{GENE-X} simulations on the $\rho_\mathrm{pol} = 0.98$ flux-surface are reported in \fref{fig:ompgradients} and compared with those used in \cite{bonanomi2024} in \tref{table:localparams}. Performing nonlinear \texttt{GENE} simulations using the \texttt{GENE-X} profiles as inputs would help assess the sensitivity of local predictions to gradient variations and quantify the role of $\bm{E}\times\bm{B}$ shear. However, this is beyond the scope of the present work.

When $S_n$ is included at $P_{\mathrm{net}} / P_{\mathrm{LH}} = 0.99$, both $P_i$ and $P_e$ are reduced yielding a better quantitative agreement with experiments. These reductions are attributed to the turbulence stabilization resulting from the flattening of the profiles with $S_n$ (see \fref{fig:densomp}). As a consequence, the particle flux is strongly reduced in the edge region (see \fref{fig:fig3}), leading to a decrease in the convective heat fluxes. While the perpendicular conductive and convective heat fluxes dominate the electron heat transport, the convective flux is the the dominant contribution to $Q_i$. The reduction of the convective heat fluxes can be visualized in the Fourier spectra depicted in \fref{fig:fluxfourier}.

\begin{figure}[h]
    \centering
    \includegraphics[scale=0.55]{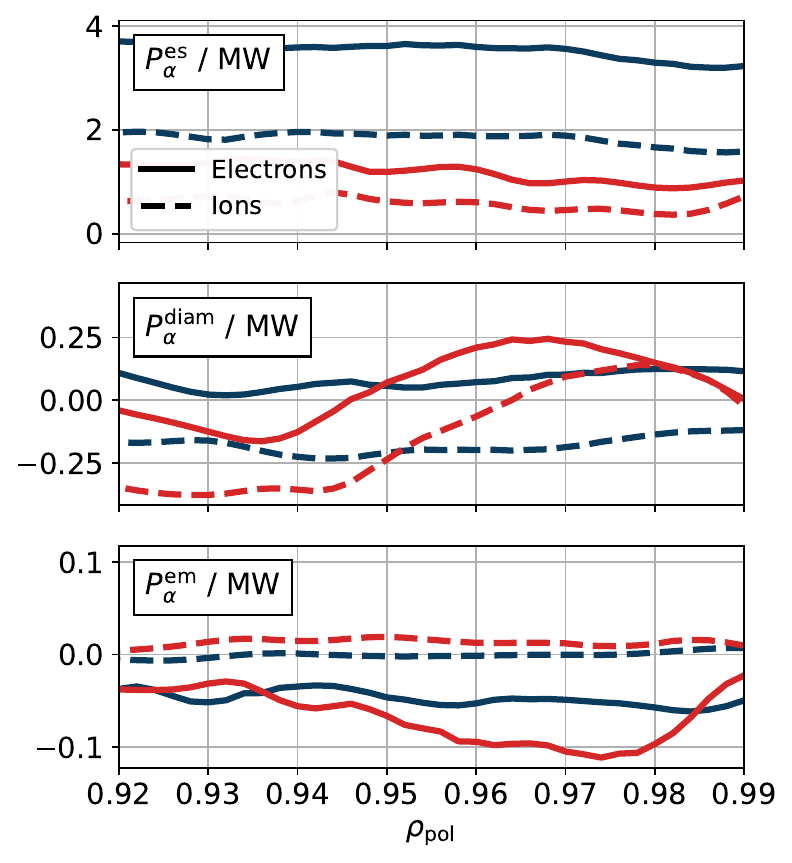}
    \caption{Radial profiles at QSS of the different contributions of $P_{\alpha}$ for electrons (solid line) and ions (dashed) at $P_{\mathrm{net}} / P_{\mathrm{LH}} = 0.99$ (from top to bottom): $P_{\mathrm{es}}$, $P_{\mathrm{diam}}$, and $P_{\mathrm{em}}$. The results with (\textcolor{tabred}{\rule{2ex}{1.5ex}}) and without (\textcolor{color4p8}{\rule{2ex}{1.5ex}}) density source $S_n$ are shown.}
    \label{fig:fluxes4p8}
\end{figure}

\Fref{fig:fluxes4p8} shows the radial profiles of the different contributions to $P_{\alpha}$ at $P_{\mathrm{net}} / P_{\mathrm{L-H}} = 0.99$ ($t = 4.8$~s). Both with and without $S_n$, $P_{\alpha}$ is dominated by turbulent $\bm{E}\times\bm{B}$ transport; however, this contribution is significantly reduced when $S_n$ is included. As a result, the diamagnetic contribution $P_{\alpha}^{\mathrm{diam}}$ becomes non-negligible. In particular, $P_i^{\mathrm{diam}} \simeq 0.2$~MW, corresponding to approximately $30\%$ of the total ion power ($P_i \simeq 0.73$~MW), whereas it remains negligible in the absence of $S_n$. The electromagnetic contribution $P_{\alpha}^{\mathrm{em}}$ is mainly carried by the electrons and is directed inwards, but its magnitude remains small compared to $P_{\alpha}^{\mathrm{diam}}$. Similar trends are observed for the particle fluxes $\Gamma_\alpha$.

These results indicate that close to the L–H transition, $P_i$ (and hence $Q_i$) arises not only from turbulent transport but also from a significant diamagnetic contribution. This highlights the need for a self-consistent treatment of turbulent and diamagnetic fluxes, which is not captured by local flux-tube approaches such as \texttt{GENE}. While the NC ion heat flux computed with \texttt{NCLASS} using experimental profiles was found negligible \cite{bonanomi2024}, such local approaches do not capture turbulence-driven modifications of the NC flux, which are retained in the present simulations \cite{frei2025}.

A summary of the ion and electron powers predicted by \texttt{GENE-X}, nonlinear flux-tube \texttt{GENE}, and the corresponding power balance estimates from \texttt{ASTRA} \cite{bonanomi2024} at $P_{\mathrm{net}} / P_{\mathrm{L–H}} = 0.99$ is provided in \tref{table:powers}. As observed, \texttt{GENE-X} predicts $P_i$ (and thus $Q_i$) that agrees with the experimental \texttt{ASTRA} estimate within $20\%$ with $S_n$, representing a better agreement than the predictions from nonlinear flux-tube \texttt{GENE} calculations. 

\begin{figure}[h]
    \centering
    \includegraphics[scale=0.55]{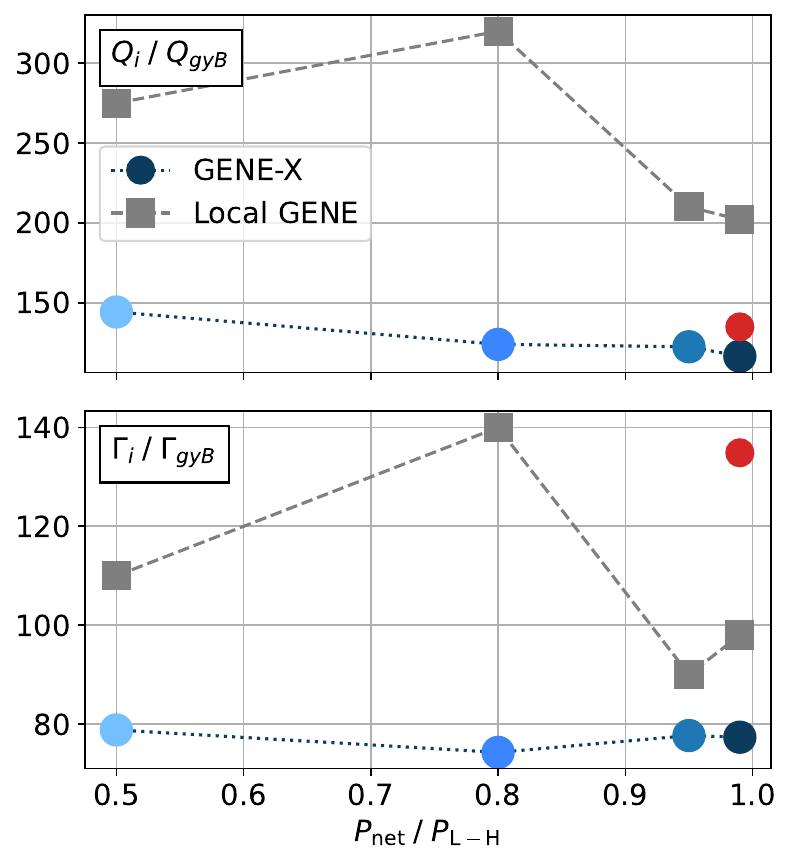}
    \caption{Total ion (top) heat and (bottom) particle fluxes, $Q_i$ and $\Gamma_i$ respectively, in gyro-Bohm units ($Q_{\mathrm{gyB}}$ and $\Gamma_{\mathrm{gyB}}$) obtained in \texttt{GENE-X} approaching the L-H transition. The \markerSquare[gray] markers indicate the results from nonlinear flux-tube \texttt{GENE} (from \cite{bonanomi2024}). See \tref{table:simsummary} for the color code. }
    \label{fig:fluxesgybohm}
\end{figure}

The increase of $P_i$ (and thus of $Q_i$) toward the L-H transition in the absence of a density source, as shown in \fref{fig:fluxesedges} and in the Fourier amplitude in \fref{fig:fluxfourier}, can be attributed to the increase of the edge ion temperature rather than to an increase in the normalized gradients, which remain similar across all time slices. In particular, when the turbulent ion heat flux $Q_i^{\mathrm{es}}$ (the dominant contribution to $Q_i$) is normalized to the gyro-Bohm heat flux, $Q_{\mathrm{gyB}} = n_e T_i c_{s} \rho_{*}^2$ (with $\rho_{*} = \rho_s / a$, $\rho_s = c_s / \Omega$, $c_s^2 = T_e / m_{i}$, and $\Omega_{i} = q_{i} B / (c m_{i})$ for hydrogen), we find $Q_i \sim Q_i^{\mathrm{es}} \sim Q_{\mathrm{gyB}}$. This gyro-Bohm scaling, indicative of local turbulent transport, is illustrated in \fref{fig:fluxesgybohm}. A similar trend is observed for the normalized particle flux, $\Gamma_i / \Gamma_{\mathrm{gyB}}$, with $\Gamma_{\mathrm{gyB}} = n_e c_{s} \rho_{*}^2$. The results with a density source are also included, but since only a single point is available, no clear gyro-Bohm scaling can be established. Nonlinear flux-tube \texttt{GENE} simulations \cite{bonanomi2024} also exhibit gyro-Bohm scaling, but the normalized ion heat fluxes exceeding those from \texttt{GENE-X}, in contrast to the trend observed in \fref{fig:fluxesedges}. This difference stems from the fact that \texttt{GENE-X} predicts higher temperatures (in particular of $T_i$) than the experimental values employed in the flux-tube \texttt{GENE} simulations.

\section{Discussion and Conclusions}

This work presents a first-principles full-$f$ GK stepwise validation of the radial electric field $E_r$ and transport approaching the \textit{experimental} L-H transition in a hydrogen AUG discharge. Using the \texttt{GENE-X} code \cite{michels2021,frei2025vspec}, simulations are performed at successive L-mode time slices with increasing heating power, spanning $0.5 \leq P_{\mathrm{net}}/P_{\mathrm{LH}} \leq 0.99$. In contrast to previous local GK studies \cite{bonanomi2024}, this approach self-consistently captures the nonlinear coupling between turbulence, equilibrium profiles, and $E_r$. An edge density source model is introduced close to the L-H transition and is found to be necessary to obtain experimentally compatible results.

Although slow profile evolution is not modeled, the OMP profiles are validated at each time slice: density profiles are underestimated without an edge density source and agreement with experiments is achieved when the source is included near the separatrix. The simulated $E_r$ well shows excellent agreement with experimental measurements approaching the L-H transition. A radial force balance analysis identifies poloidal flows as the dominant contribution to the $E_r$ well. While consistent with NC predictions in the edge region, large deviations close to the separatrix indicate the importance of nonlinear turbulence-flow interactions.

Edge turbulence is characterized by competing eDWs and TEMs, in qualitative agreement with previous finding from linear and local GK analysis \cite{bonanomi2024}. We find that the turbulence properties (frequency and wavenumber relation) remain unchanged between the different time slices. Nonetheless, the inclusion of the density source reduces density gradients, leading to strong turbulence stabilization and reduced turbulent transport. While $\bm{E}\times\bm{B}$ shearing rates comparable to those inferred from experiments in \cite{bonanomi2024} are obtained (see \tref{table:localparams}), their role in turbulence mitigation (together with $\beta_e$ destabilization) in the present cases remains to be assessed and can only be addressed using a local approach.

The heat transport across the separatrix measured in the simulations reproduces key experimental trends, including larger electron heat flux associated with ECRH power and increasing electron and ion heat fluxes as the L-H transition is approached. In the absence of a density source, edge ion heat and particle fluxes follow a Gyro-Bohm scaling dominated by nonlinear $\bm{E}\times\bm{B}$ advection but overestimate the total power crossing the separatrix. Including the density source reduces the $\bm{E}\times\bm{B}$ flux though turbulence stabilization and enhances the ion diamagnetic contribution, yielding an ion heat flux in close agreement with experiments close the L-H transition and improving the previous local nonlinear calculations \cite{bonanomi2024}. This demonstrates the non-negligible role of ion diamagnetic transport in $Q_i$ in the edge near the L-H transition. On the other hand, electromagnetic contributions remain negligible in all cases.

The present full-$f$ simulations allows to identify several fundamental limitations of local GK simulations \cite{bonanomi2024} although the latter approach reproduce turbulent flux levels close to experiments. First, the evolution of the $E_r$ well (and its associated shear) is governed by nonlinear turbulence-mean poloidal flow interactions, which is absent when $E_r$ is externally imposed. Second, predicting experimentally compatible \textit{total} ion transport near the L-H transition requires a consistent treatment of turbulent and diamagnetic fluxes. While turbulent fluxes can be varied within experimental uncertainties in local calculations, accurate prediction of diamagnetic fluxes requires a global full-$f$ formalism, as long wavelength (NC) components can be modified by shorter wavelength turbulence \cite{frei2025}. Third, the nonlinear edge dynamics is highly sensitive to the properties of the edge density source, highlighting the need for accurate particle source modeling to correctly predict edge transport, $E_r$, and ion heat fluxes. Beyond these physical insights, the present simulations provide also a valuable dataset for the development of reduced transport models. In particular, turbulent flux spectra from full-$f$ GK simulations (such as the ones shown in \sref{sec:appendixc}) can inform nonlinear saturation models in reduced quasi-linear frameworks such as \texttt{TGLF} \cite{staebler2007}.

Finally, although slow profile evolution across successive ECRH power steps is not directly simulation, the present stepwise validation demonstrates that experimentally relevant pre–L-H transition conditions, including $E_r$ structure and ion heat fluxes, can be reproduced using the full-$f$ GK code \texttt{GENE-X} at AUG. The physics fidelity of the model will be further enhanced by adding self-consistent neutral gas dynamics and FLR effects. This work provides a solid foundation for future flux-driven, dynamically resolved GK studies of the L-H transition \cite{zholobenko2026}.

\section*{Acknowledgement}

The authors are grateful to G.D. Conway (and U. Plank) for the measurements of the radial electric field profiles which have been used in this work for comparisons with the simulations.

The simulations presented in this work were performed on the Phase 1 partition of the SuperMUC-NG system at the Leibniz Supercomputing Centre (LRZ), under Project ID \texttt{pn39do}. 

This work has been carried out within the framework of the EUROfusion Consortium, funded by the European Union via the Euratom Research and Training Programme (Grant Agreement No 101052200 — EUROfusion). Views and opinions expressed are however those of the author(s) only and do not necessarily reflect those of the European Union or the European Commission. Neither the European Union nor the European Commission can be held responsible for them.

\appendix

\section{Numerical Aspects}
\label{sec:appendixa}

The normalization used in \texttt{GENE-X} for the velocity-space spectral approach is detailed in \cite{frei2025vspec}. All simulations use the following reference quantities: a length $L_{\mathrm{ref}} = 1.65$~m, a magnetic field $B_{\mathrm{ref}} = 2.5$~T, a temperature $T_{\mathrm{ref}} = 100$~eV, a density $n_{\mathrm{ref}} = 10^{19}$~m$^{-3}$, and a reference mass $m_{\mathrm{ref}} = m_p$ ( $m_p$ is the proton mass). These values are representative of typical AUG L-mode conditions.

The configuration space in \texttt{GENE-X} is discretized using the FCI approach \cite{stegmeir2019}. A total of $32$ poloidal planes is employed to represent the toroidal direction ($\phi$ direction). Within each poloidal plane, a Cartesian grid with a uniform spacing of $\Delta RZ = 1.43$~mm is used. This corresponds to approximately $1.8 \rho_i$ at the separatrix for $T_i \sim 150$~eV (see \fref{fig:tempomp}). This resolution is sufficient to capture fluctuations with $k_y \rho_i \lesssim 1$, consistent with the long-wavelength ordering of the \texttt{GENE-X} physical model (see \sref{sec:sec3}). This constitutes the baseline resolution used in the present simulations. A resolution scan performed at $t = 3.8$~s using a finer grid spacing of $\Delta RZ = 0.99$~mm ($\sim 1.3 \rho_i$ at the separatrix) shows only minor differences in physical observations.

In velocity space, we use $6$ parallel (Hermite in $v_\parallel$) and $4$ perpendicular (Laguerre in $\mu$) spectral coefficients, with reference scaling temperatures $\tau_i = 2.75 T_{\mathrm{ref}}$ and $\tau_e = 5 T_{\mathrm{ref}}$. This resolution choice is motivated by previous L-mode validation studies \cite{frei2025} and is found to adequately resolve the relevant dynamics. Increasing the velocity-space resolution to $12$ Hermite and $6$ Laguerre spectral coefficients at $t = 3.8$~s confirms the convergence.

An explicit fourth-order Runge-Kutta scheme is used for time integration with a normalized time step of $\Delta t = 3 \times 10^{-4}$.
The simulations were performed on the SuperMUC-NG Phase 1 supercomputer at LRZ, equipped with Intel Xeon Platinum~8174 CPUs ($48$~cores per node) across $6480$ nodes. Each simulation was executed using $64$~nodes ($1$ MPI rank per node). Reaching a quasi-steady state without the density source $S_n$ requires $1.6 \cdot 10^5$~time steps, corresponding to a computational cost of approximately $0.6$~MCPUh per simulation and, therefore, a total of $2.5$~MPCUh for all simulations without the density source $S_n$. Including $S_n$ increased the required simulation length to $3\cdot 10^5$~time steps, yielding a computational cost of approximately $1.15$~MCPUh. Overall, the simulations presented in this work required a total computational cost of approximately $3.6$~MCPUh. 

\section{Impact of Density Source Properties}
\label{sec:appendixb}

The localized density source $S_n$ introduced in \sref{subsec3.1} depends on fixed parameters such as amplitude, radial width, and position. In the present simulations, these parameters are adjusted to reproduce the experimental density profiles near the separatrix. However, the spatial distribution of the density source due to neutral ionization processes is considerably more complex than the simplified model adopted here. It is therefore of interest to assess how variations in the source parameters affect $E_r$.

In addition to the reference source model $S_n$ introduced in \sref{subsec3.2}, we consider two modified density source models, $S_n^{(1)}$ and $S_n^{(2)}$. Both $S_n^{(1)}$ and $S_n^{(2)}$ have the same amplitude as $S_n$ ($1.6 \times 10^{22}\,\mathrm{s}^{-1}$), but are radially shifted inward into the edge region to $\rho_S = 0.98$ with a broader radial extent, $L_S = 0.01$. In addition, $S_n^{(2)}$ neglects the energy term in \eqref{eq:sourcevspec}, which leads to a non-vanishing energy source term $\mathbb{S}_{\epsilon \alpha} = 3 \mathbb{S}_{n_\alpha} T_S / 2$ in the energy balance equation (see \eqref{eq:steadystateheat2}). As a consequence, $S_n^{(2)}$ effectively increases the surface-integrated energy flux $Q_{\mathrm{tot}}$ (see \eqref{eq:steadystateheat2}) by approximately $3$~MW at the source location, with contributions of about $1$~MW from ions and $1.9$~MW from electrons. In contrast, this energy source term associated with $S_n$ and $S_n^{(1)}$ vanishes, $\mathbb{S}_{\epsilon \alpha} = 0$. 

\begin{figure}[h]
    \centering
    \includegraphics[scale=0.55]{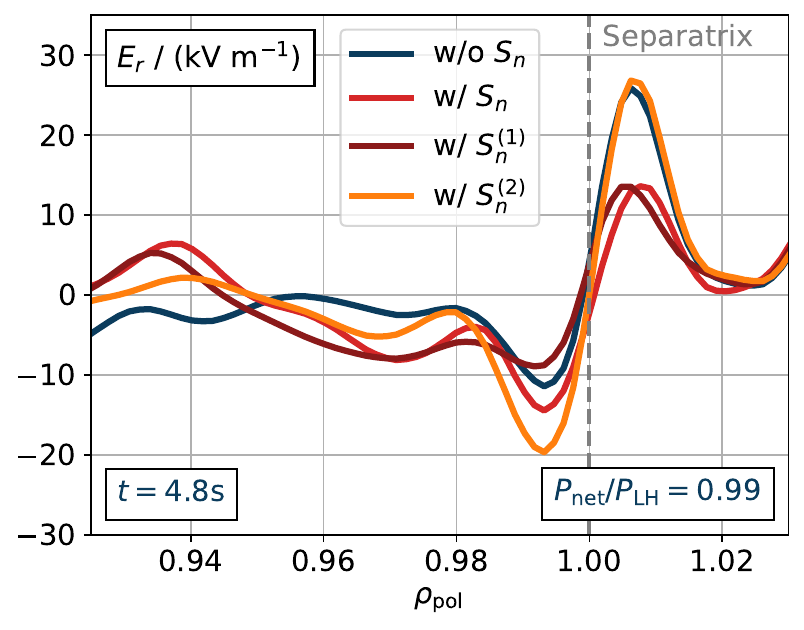}
    \caption{OMP $E_r$ profiles at $t=4.8$~s (near the L–H transition) obtained with different density sources models (\textcolor{tabred}{\rule{2ex}{1.5ex}}): $S_n$ (solid line), $S_n^{(1)}$ (brown solid line), $S_n^{(2)}$ (orange solid line). The case without a density source (w/o $S_n$) is shown by the solid (\textcolor{color4p8}{\rule{2ex}{1.5ex}}) line. The separatrix is marked by the vertical dashed gray line. }
    \label{fig:ersourcs}
\end{figure}

\begin{figure}[h]
    \centering
    \includegraphics[scale=0.55]{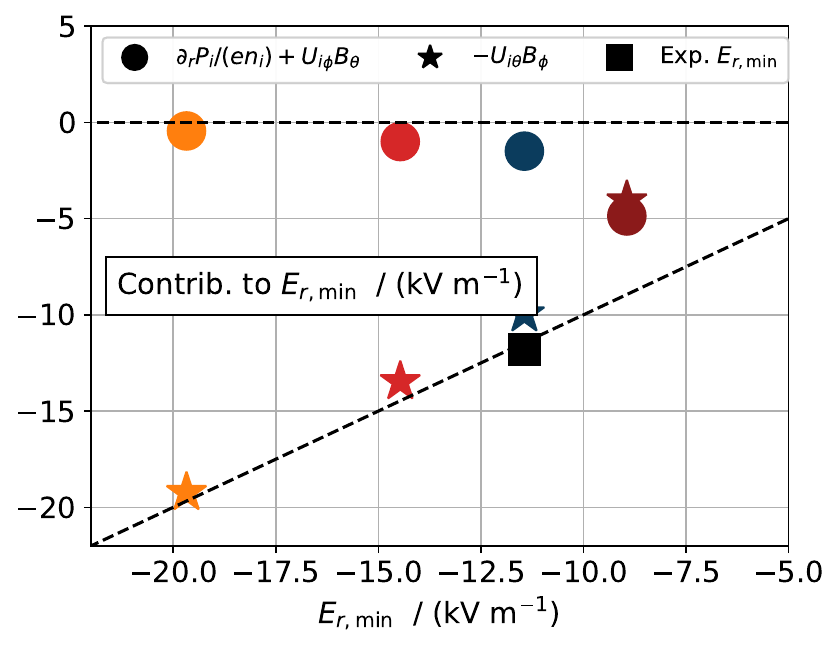}
    \caption{Same as \fref{fig:ermin} at $t = 4.8$~s ($P_{\mathrm{net}}/P_{\mathrm{LH}} = 0.99$) including the density source models $S_n^{(1)}$ and $S_n^{(2)}$. The same colors as in \fref{fig:ersourcs} are used.}
    \label{fig:erminsources}
\end{figure}

We perform two simulations at $t = 4.8$~s with $S_n^{(1)}$ and $S_n^{(2)}$ and compare the resulting OMP $E_r$ profiles in \fref{fig:ersourcs}. With $S_n^{(1)}$, the $E_r$ well is wider and shallower than in the reference case $S_n$, with a reduced minimum value $E_{r,\min}$. The density profiles remain very similar (not shown), whereas the temperature profiles are flattened further inside the plasma due to the local temperature reduction associated with the source. Contrary to the reference case with $S_n$, the diamagnetic and toroidal flow contributions have a similar contributions to $E_{r,\min}$ as the poloidal flows, as shown in \fref{fig:erminsources}. However, these poloidal flows are much weaker than with $S_n$, resulting in a shallower $E_r$ well. 

In contrast, $S_n^{(2)}$ produces a deeper and narrower $E_r$ well. In this case and similarly than in \sref{sec:sec6}, poloidal flows dominate the $E_r$ well and are found much stronger as shown in \fref{fig:erminsources}. In this case, $E_{r,\min}$ approaches typical H-mode values observed at AUG \cite{sauter2011,viezzer2013,plank2023}. We remark that the temperature profiles are similar to those obtained without a density source as the increase in heat flux associated with the finite energy source term $\mathbb{S}_{\epsilon \alpha}$ compensates for the local temperature reduction induced by $S_n^{(2)}$. This yields larger temperature gradients near the separatrix. On the other hand, the density profile remains very similar. This local increase of the temperature gradients (with similar density gradients) provide an additional turbulence drive, resulting in an enhanced turbulence-mean flow interaction yielding the stronger poloidal flow observed in this case. GAM activity is also found to be stronger with $S_n^{(2)}$. However, no clear interactions with turbulence is found.

\section{Flux Fourier Analysis} \label{sec:appendixc}
In this section, we present the Fourier spectra of the turbulent fluxes associated with the $\bm{E} \times \bm{B}$ radial transport of particles and heat. The turbulent $\bm{E} \times \bm{B}$ heat flux is decomposed into a convective part, $Q_{\alpha}^{\mathrm{conv}}$ (associated with density fluctuations), and parallel and perpendicular conductive parts, $Q_{\parallel \alpha}^{\mathrm{cond}}$ and $Q_{\perp \alpha}^{\mathrm{cond}}$ (associated with parallel and perpendicular temperature fluctuations). Their definitions can be found in \cite{frei2025}. 

\begin{figure*}[h]
    \centering
    \includegraphics[scale=0.55]{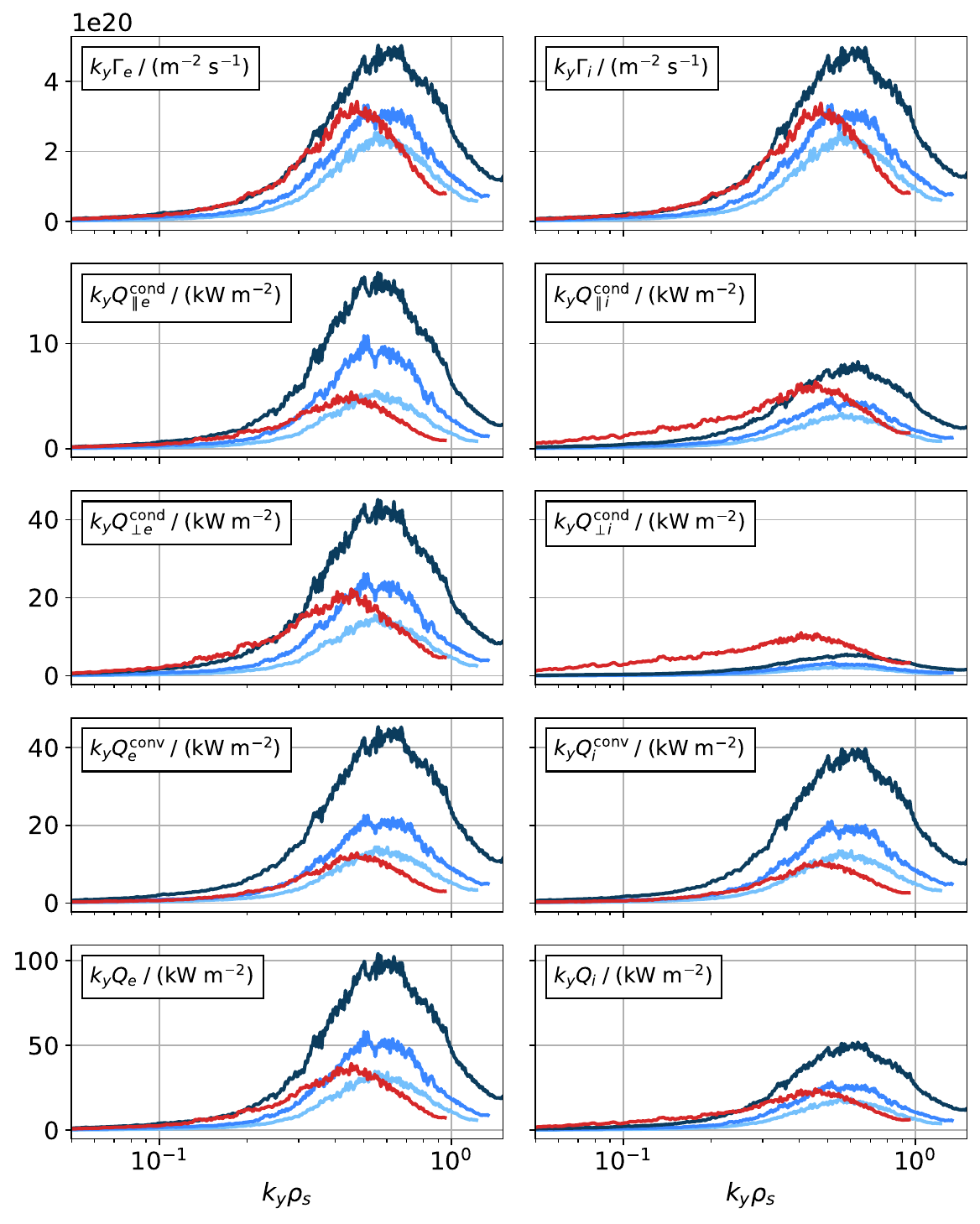}
    \caption{Fourier spectra of the electron (left column) and ion (right column) fluxes, evaluated at the $\rho_{\mathrm{pol}} = 0.98$ flux surface, as a function of the normalized binormal wavenumber $k_y \rho_s$. The particle, conductive parallel and perpendicular, convective and total fluxes are shown from top to bottom, respectively. The results from the time slices $t=2.7$~s, $3.8$~s, and $4.8$~s with and without $S_n$ are shown (see \tref{table:simsummary}).}
    \label{fig:fluxfourier}
\end{figure*}

We show the results in \fref{fig:fluxfourier} of the time slices at $t = 2.7$~s ($P_{\mathrm{net}}/P_{\mathrm{LH}} = 0.5$), $t = 3.8$~s ($P_{\mathrm{net}}/P_{\mathrm{LH}} = 0.85$), and $t = 4.8$~s ($P_{\mathrm{net}}/P_{\mathrm{LH}} = 0.99$), with and without $S_n$. As the L–H transition is approached, the increase in heat fluxes is clearly reflected in the larger amplitudes in the spectra, peaking near $k_y \rho_s \sim 0.5$. However, the shape of the spectra remains qualitatively similar. The stabilizing effect of the density source $S_n$ is also visible.

\section{Normalized Gradients and Local Plasma Parameters}
\label{sec:appendixd}

\begin{figure}[h]
    \centering
\includegraphics[scale=0.57]{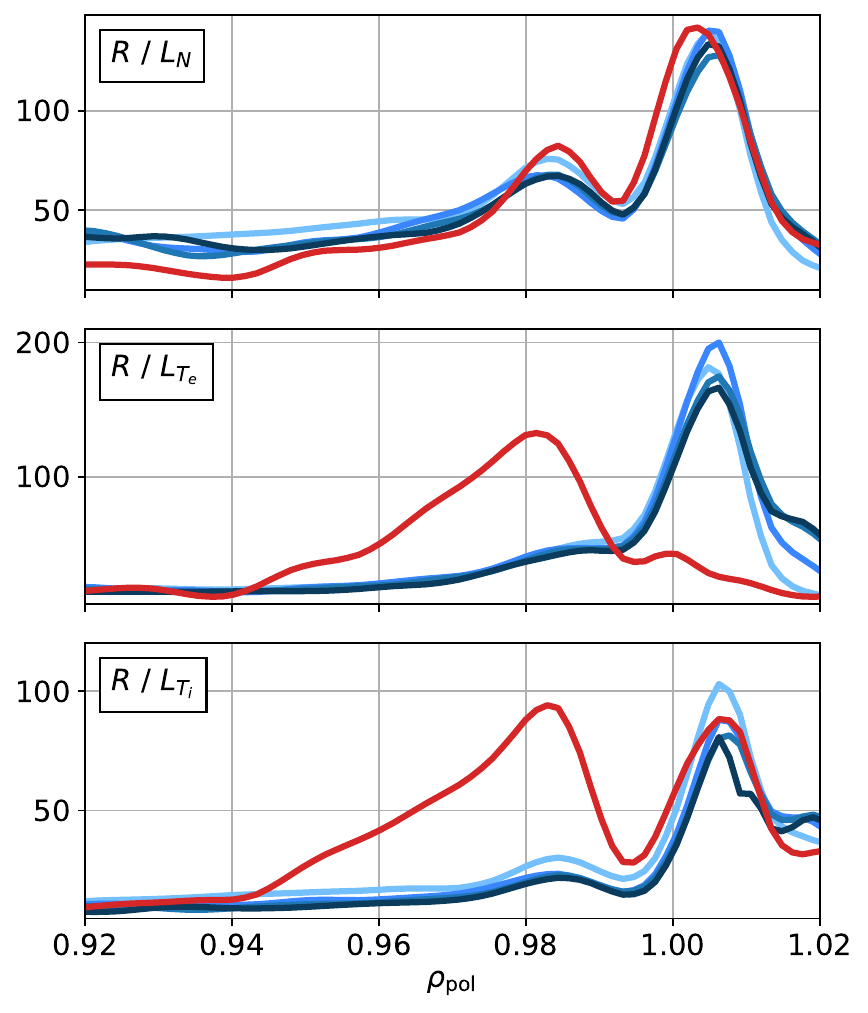}
    \caption{Normalized density (top) and temperature (bottom) normalized gradient, $R / L_f$, associated with the OMP profiles (see \fref{fig:densomp} and \fref{fig:tempomp}) for the different times slices (see \tref{table:simsummary}).}    
    \label{fig:ompgradients}
\end{figure}

From the OMP density and temperatures profiles displayed in \fref{fig:densomp} and \fref{fig:tempomp}, the normalized gradient scale length, $R / L_f = - R  \nabla f / f$ can be computed at the OMP and are displayed in \fref{fig:ompgradients}. The normalized gradient scale length remains similar for the different time slices without $S_n$. This is in contrast to the gradual increase observed from the experimental profiles, as presented in \ref{table:localparams} where the values of $R / L_f$ are reported on the $\rho_{\mathrm{pol}} = 0.98$ flux-surface. This contrasting evolution of $R / L_f$ is partially due to the increase of the SOL temperatures simulated in \texttt{GENE-X} near the separatrix yielding to a weaker increase of the edge gradients.

\Tref{table:localparams} reports the main local plasma parameters calculated on the $\rho_\mathrm{pol} = 0.98$ flux-surface from the \texttt{GENE-X} simulations for the different time slices. There parameters are also compared with the ones used in the local flux-tube \texttt{GENE} calculations from \cite{bonanomi2024}. 

\begin{table*}[t]
\centering
\begin{tabular}{cccccc}
 $t$ / s & $P_{\mathrm{net}} / P_{\mathrm{LH}}$ & $R / L_N$  & $R / L_{T_i}$ & $R / L_{T_e}$ & $\beta_e$ ($10^{-3}$)\\
\hline
\hline
$2.7$ &  $0.5$ & 45 / \textbf{\textcolor{color2p7}{70}} & $21$ / \textbf{\textcolor{color2p7}{66}}  & $50$ / \textbf{\textcolor{color2p7}{37}}  & $0.14$ / \textbf{\textcolor{color2p7}{0.11}}  \\
$3.8$ &  $0.8$ & $38$ / \textbf{\textcolor{color3p8}{66}} & $29$ / \textbf{\textcolor{color3p8}{21}}  & $35$ / \textbf{\textcolor{color3p8}{40}} & $0.185$ / \textbf{\textcolor{color3p8}{0.14}}  \\
$4.5$ &  $0.95$ & $52$ / \textbf{\textcolor{color4p5}{62}} & $40$ / \textbf{\textcolor{color4p5}{20}}  & $52$ / \textbf{\textcolor{color4p5}{37}} & $0.25$ / \textbf{\textcolor{color4p5}{0.18}} \\
$4.8$ &  $0.99$ & $52$ / \textbf{\textcolor{color4p8}{62}}  & $43$ / \textbf{\textcolor{color4p8}{19}}  & $53$ / \textbf{\textcolor{color4p8}{36}}  & $0.3$ / \textbf{\textcolor{color4p8}{0.21}}  \\
$4.8$ &  $0.99$ & $52$ / 
\textbf{\textcolor{tabred}{69}}  & $43$ / \textbf{\textcolor{tabred}{87}}  & $53$ / \textbf{\textcolor{tabred}{131}}  & $0.3$ / \textbf{\textcolor{tabred}{0.15}} \\
\hline
\end{tabular}
\caption{Local plasma parameters used in the nonlinear \texttt{GENE} calculations (back text) from \cite{bonanomi2024} and calculated from the \texttt{GENE-X} simulations (colored text). Here, $\beta_e  = 8 \pi P_e / B^2$ is the electron plasma beta. The quantities are evaluated at the OMP on the $\rho_\mathrm{pol} = 0.98$ flux-surface.}
\label{table:localparams}
\end{table*}


\section*{References}
\bibliography{library}

\end{document}